\documentclass[sigconf]{acmart}

\usepackage[english]{babel}
\usepackage{blindtext}

\usepackage{algorithm} 
\usepackage{algpseudocode} 
\usepackage{amsmath} 
\usepackage{graphicx}
\usepackage{tabularx}
\usepackage{subcaption}
\usepackage{enumitem}

\renewcommand\footnotetextcopyrightpermission[1]{} 
\setcopyright{none}

\acmDOI{}

\acmISBN{}

\acmConference[Preprint]

\acmPrice{}

\begin{document}
\title{FIKIT: Priority-Based Real-time GPU Multi-tasking Scheduling with Kernel Identification}


\author{Wenqing Wu}
\affiliation{%
  \institution{Department of Computer Science, University of Chicago}
  \city{Chicago} 
  \state{USA} 
}
\email{wenqingw@uchicago.edu}

\renewcommand{\shortauthors}{Wu, Wenqing}

\begin{sloppypar}
\begin{abstract}
    Highly parallelized workloads like machine learning training, inferences and general HPC tasks are greatly accelerated using GPU devices. In a cloud computing cluster, serving a GPU's computation power through multi-tasks sharing is highly demanded since there are always more task requests than the number of GPU available. Existing GPU sharing solutions focus on reducing task-level waiting time or task-level switching costs when multiple jobs competing for a single GPU. Non-stopped computation requests come with different priorities, having non-symmetric impact on QoS for sharing a GPU device. Existing work missed the kernel-level optimization opportunity brought by this setting. To address this problem, we present a novel kernel-level scheduling strategy called FIKIT: Filling Inter-kernel Idle Time. FIKIT incorporates task-level priority information, fine-grained kernel identification, and kernel measurement, allowing low priorities task's execution during high priority task's inter-kernel idle time. Thereby, filling the GPU's device runtime fully, and reduce overall GPU sharing impact to cloud services. Across a set of ML models, the FIKIT based inference system accelerated high priority tasks by 1.32 to 16.41 times compared to the JCT in GPU sharing mode, and more than half of the cases are accelerated by more than 3.4 times. Alternatively, under preemptive sharing, the low-priority tasks have a comparable to default GPU sharing mode JCT, with a 0.86 to 1 times ratio. We further limit the kernel measurement and runtime fine-grained kernel scheduling overhead to less than 5\%. 
\end{abstract}

\maketitle

\section{Introduction}
In multitasking and containerized cloud computing environments (referred as “cloud computing environment”), tasks from different scenario needs to be served with different priorities. For example, real-time recommender and AI Inferencing services need prioritized access to GPU resources for low latency and high throughput service-level objectives. Time insensitive data analytics tasks, however, can share resources at the background of real-time services, scavenging real-time services' idling resource. There has been much research on improving concurrent efficiency in multi-level priority GPU scheduling \cite{10,11,16,18,19,21,27,28,31,32,33,34}, and previous works \cite{10,13,30} on tasks scheduling within a single GPU (referred as "Intra-GPU Scheduling" in this paper).

This paper focuses on Intra-GPU scheduling, exploring a scheduling system with both GPU task preemption capability and multitask concurrency. It can interleave high and low priority tasks, while minimizing performance impact on high priority tasks.

\begin{figure}[tp]
	\centering
	\includegraphics[width=\columnwidth]{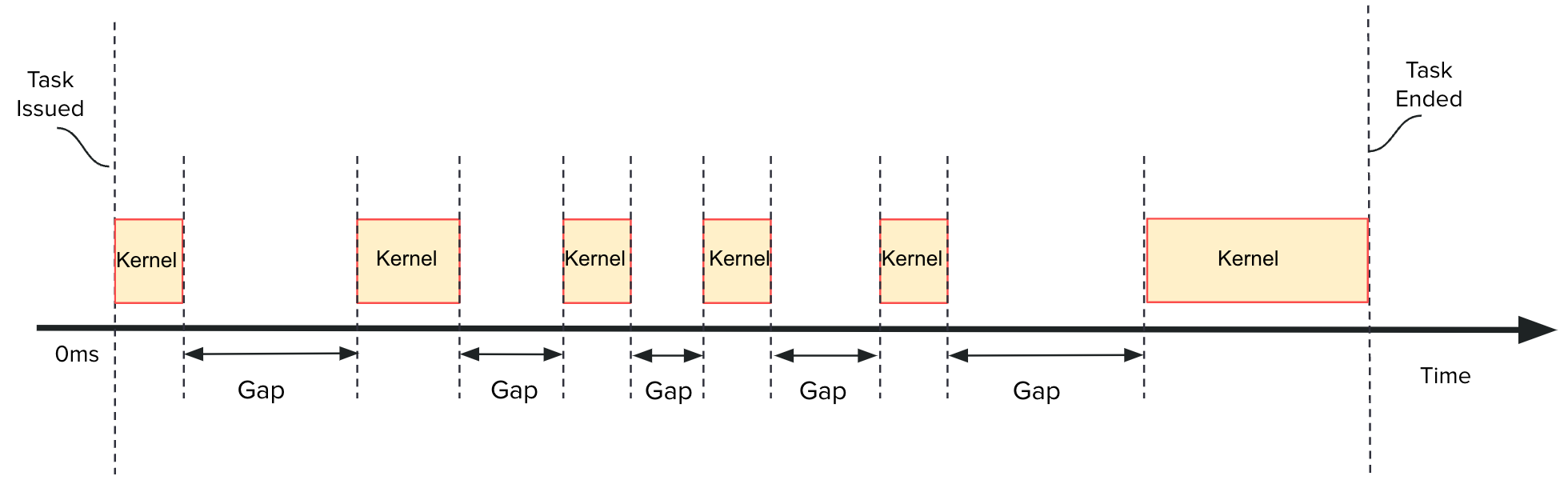}
	\caption{A GPU task has gaps between kernels.}
	\label{fig:kernelDemonstration}
\end{figure}

As shown in Figure~\ref{fig:kernelDemonstration}, when a task independently uses a GPU, inference tasks have gaps between executing kernels on the GPU. These gaps arise from several factors, such as CPU/Memory/Launch overhead\cite{43,44,45}, launch latency \cite{43,44,45}, insufficient compute intensity \cite{46,47}, load imbalance, and the presence of GPU blocking operations \cite{48}. Several studies have focused on reducing this gap, such as: lowering overhead and latency, increasing compute intensity, reducing GPU blocking operations, etc. These optimizations are application specific, not universal, or generalizable, and eliminating the inter-kernel gaps is unrealistic. In the cloud computing environment, inference tasks are often executed in short durations, and some customers simply do not care about optimizing the execution efficiency of inference tasks, even if the inference tasks they run contains "large" Gaps. In summary, GPU tasks with low GPU saturation and large gaps is prominent in cloud computing environments. Our new approach to this problem taking advantages of these “large” gaps instead of eliminating them: inserting kernels that execute low priority tasks during high priority task's inter-kernel idle time. This method is both generalizable and effective.

Based on this idea, we propose a new Filling Inter-kernel Idle Time scheduling strategy (denoted as FIKIT). It can execute low priorities task's GPU kernel during high priority task's inter-kernel idle time (Hereafter denoted as “gap”). Thereby, filling the GPU's device runtime fully, maximizing the GPU's computing capacity.

To implement the FIKIT scheduling strategy and make full use of inter-kernel GPU idle gaps, it is necessary to solve several key challenges that are not touched by previous works:
1) Under the closed-sourced Nvidia GPU driver environment, how to accurately identify each kernel called by the task using GPU? It also needs to measure the execute time of each kernel, and task's inter-kernel idle gaps.
2) Based on kernel execution time and gaps, how to carry out real-time kernel-level scheduling and keep the GPU overhead of scheduling less than 5\%?

For intra-GPU scheduling, Nvidia provides several solutions. Its GPU are equipped with the default sharing mode, which inter-kernel gaps are reduced through interleaving kernels from different tasks. It, however, favors a kernel-level fair scheduling, not proactively filling gaps, so large gaps are still widely present and are wasted. The more pressing issue is that it is based on a FIFO device queue, it cannot preempt low-priority tasks which results in the job completion time (JCT) of high-priority tasks running in this mode being several times higher than that of running exclusively. In Nvidia GPU's exclusive mode, it only supports launching a single task and does not have the capability for multi-task scheduling. Nvidia MPS \cite{4} enables some multi-tasks concurrency by merging GPU execution contexts of sharing processes into one. However, MPS still has limitations. MPS does not support priority-based preemption of lower priority processes to free up GPUs for high-priority processes. NVIDIA's vGPU\cite{5}, vCS\cite{6} and vDWS\cite{7} technologies all try to deal with GPU sharing and scheduling problems, but the lack of preemptive mechanisms leads to limited efficiency and flexibility.

Previous work like TimeGraph\cite{10}, REEF\cite{30} and Gemini\cite{13} have implemented kernel-level schedulers, but they did not propose the inter-kernel gap filling mechanism. TimeGraph\cite{10} and REEF\cite{30} implemented schedulers based on different open source drivers, modifying open source code to achieve richer mechanisms to orchestrate GPU tasks, without needing to use a mechanism like FIKIT. However, these schedulers are difficult to be widely adopted in cloud computing environments as open-source driver and services are hardly available. Gemini\cite{13} adopts methods such as "incidentally" inserting monitoring events when the application evokes synchronization events to reduce the overhead caused by CUDA hooks. It did not implement a fine-grained inter-kernel gap filling mechanism. Gemini would not be able to perform such scheduling as it cannot identify each kernel function.

We developed a kernel-level scheduling framework, FIKIT framework, which consists of a recompiled pytorch/tensorflow library, a hook client per service and the FIKIT scheduler application. The recompiled pytorch/tensorflow is deployed in cloud environments once and used by all services, replacing the native pytorch/tensorflow library. The hook client is responsible for collecting the kernel-level statistics during the profiling phase and constructing the kernel id to identify each kernel in real-time during the scheduling phase. The FIKIT scheduler is responsible for scheduling the preemption of different priority tasks and implementing the FIKIT scheduling algorithm.

Overall, our paper made the following contributions:
\begin{enumerate}
	\item A novel idea is proposed: Instead of eliminating inter-kernel idle gaps, we fully utilize the "larger" Gaps to implement a new kernel-level scheduling mechanism. In a cloud computing environment, we use the gaps of a running GPU task to run the kernel of other GPU tasks i.e., FIKIT scheduling algorithm. 
	\item A new and efficient method for identifying GPU kernels and measuring kernel execution time and inter-kernel idle time (gap) has been implemented. The new method enables offline measurement of GPU kernel function execution time to be applied in an online real-time manner, thereby greatly improving the efficiency of FIKIT scheduling. Its performance overhead on GPU task running can be almost ignored (compared with the default environment, the change of job completion time (JCT) is between -2\% to 2\%).
	\item Based on the real-time identification and efficient measurement of GPU kernels, we divide GPU tasks into two stages: measurement and sharing, which avoids the high GPU performance overhead caused by introducing kernel-level measurement, limiting FIKIT overhead under 5\%.
\end{enumerate}

\section{Background And Motivation}
\subsection{Background}
This paper hopes to design a more efficient and flexible Intra-GPU Scheduling system to ensure short JCT of high-priority tasks and concurrently execute as much  low-priority tasks as possible to maximize the utilization of GPU computation time. The following section will introduce the relevant technology and terms involved in this paper, and their relationship with the algorithms and strategies proposed in this paper

\textbf{Time slicing }is a technique to enable concurrent multi-tasking on a single GPU by dividing time into a series of time slices and allocating each time slice to  different tasks, and thus sharing the hardware resources. Time slicing is the foundational technique of this paper. There have been many studies in this field, but some studies(e.g. Gemini\cite{13}) tend to focus on fair sharing of GPU resources, unable to guarantee the QoS of JCT for important tasks. Other approaches (i.e. REEF\cite{30},TimeGrapsh\cite{10}) operate at the driver level, relying on open-source driver code or cracking closed-source driver code to achieve kernel-level scheduling. This lacks commercial viability.

\textbf{MPS (Multi-Process Service)}
aims to improve the performance of multi-process applications by lowering the overhead caused by context switching. MPS enables a form of time-slicing sharing by allowing all CUDA processes to share a single hardware context. It interferes with the time-slicing mechanism proposed in this paper. Therefore, the scheduling design in this paper does not apply to GPUs with MPS technology enabled.

 \textbf{MIG (Multi-Instance GPU)} can partition a single GPU at the physical level into multiple smaller GPU instances, each with its own memory, compute units and other hardware resources. This spatial multiplexing approach enforces real hardware isolation and fair resource allocation. MIG partitioning can coexist with time-slicing schedulings. the scheduling design of this paper can apply to a single GPU instance under MIG partitioning.

\textbf{A Single GPU} refers to a single GPU device or a GPU instance within a MIG enabled hardware, such as: a single GeForce RTX 3090, GeForce RTX 4080Ti and other GPU, or an GPU instance within NVIDIA H100, A100 GPUs. 

\textbf{Job Completion Time (JCT)} describes the total amount of time a job takes from start to finish. This typically includes wait time, actual execution time, and possibly other delays (e.g., I/O operations, network latency, etc.)

\textbf{Multitasking environment} refers to the scenario of concurrently running multiple training tasks or inference tasks on a single GPU. This paper focus on multitasking with different task priorities and are also compatible with all tasks having the same priorities (degraded to ordinary sharing mode). In a multitasking environment, different tasks have heterogeneous requirements for computing resources and memory bandwidth. There have been many published papers\cite{10,11,16,18,19,21,27,28,31,32,33,34} on how to reasonably schedule multiple tasks to run concurrently on one or multiple GPUs. This paper assumes that the current GPU resources have met the concurrency requirements for running multiple tasks at the same time.

\textbf{Container-based GPU multitasking} can improve resource utilization. This work is applicable to the containerized GPU multitasking environment. There have been extensive studies\cite{11,27,34} on how to plan and schedule GPU resources under containerized environment based on task-level resource requirements. However, task-level resource allocation under containerized environment is not the focus of this paper. This paper assumes GPU resources have met the concurrent running needs of multiple tasks within containers, and does not explore the details of scheduling tasks to a GPU device.

\textbf{Cloud computing environment} allows dynamic resource scheduling allocation and elastic adjustment of resources based on task demand. However, this work does not explore resource scheduling strategies in cloud computing environments or in large-scale cluster management systems. This work makes full use of the characteristics of the cloud computing environment, which usually makes high volume of repeated calls to a hosted cloud services, and optimizes the GPU multitasking scheduling.

\subsection{Motivation}
Our first point of view: In the cloud computing environment, the ideal GPU concurrency mode should have both task preemption mechanism and multitasking concurrent execution. It allows important high-priority tasks to run faster than the shared mode, ensures the low JCT value of important tasks, and performs as many low-priority tasks as possible. Through this preemptive multitasking concurrency mode, the GPU can further improve the overall multitasking throughput and achieve more efficient resource utilization while ensuring the execution of high-priority tasks.

Our second point of view: In the cloud computing environment, a considerable number of cloud services will always have "large" inter-kernel gaps. By utilizing these "large" gaps to execute other tasks' kernel, the GPU can be made busier.

\begin{figure}[tp]
    \centering
    \includegraphics[width=\columnwidth]{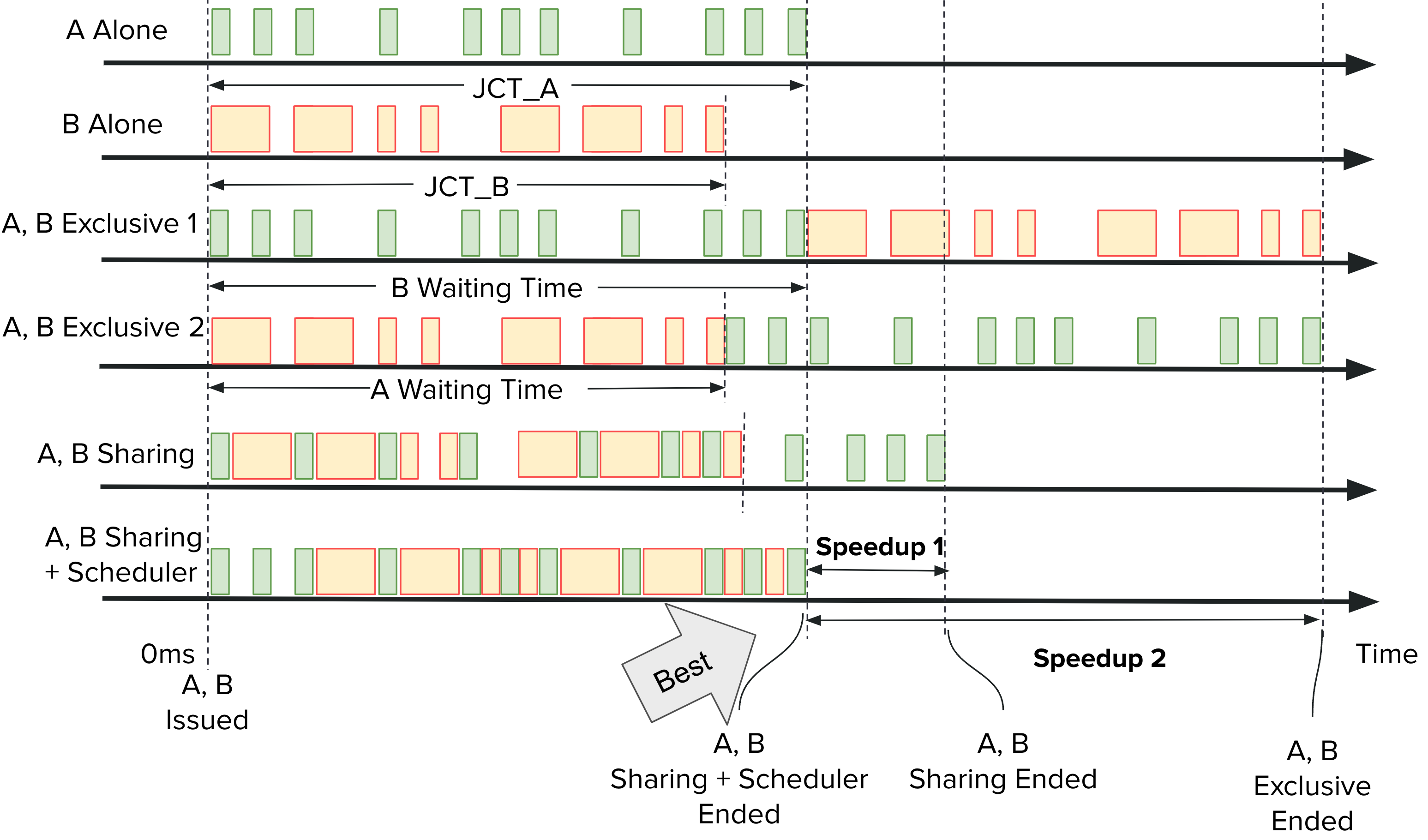}
    \caption{Using an idealized two tasks sharing case study, we demonstrated how FIKIT works compared to Nvidia GPU sharing and exclusive scheduling.}
    \label{fig:caseStudy}
\end{figure}

For example, GPU tasks A and B are running concurrently. We assumed that task A is a time-sensitive high-priority task, and task B is a low-priority task that is allowed to have some delay. The following evaluation metrics are designed.
\begin{enumerate}
    \item Assuming tasks A and B execute concurrently, under scheduling mode i, the JCT values for the two tasks are $JCT_A$ and $JCT_B$ respectively when it occupies the GPU exclusively, and the JCT values of the two tasks in scheduling mode i are $JCT_{A_i}$ and $JCT_{B_i}$, respectively, $JCT_{A_i}$ and $JCT_{B_i}$ should be less than $JCT_A + JCT_B$; the JCT ratio of high priority task A under scheduling mode i versus exclusive mode, $JCT_{A_i} / JCT_A$ is an important evaluation index. The closer it is to 1 the better. The JCT value, $JCT_{B_i}$, for low priority task B must be stable and predictable, i.e. its value needs to stabilize within a certain range.
    \item Assuming task B runs continuously, and task A is inserted to run. At this time, the JCT of task A is $JCT_{\text{A insert}}$. The closer the value of $JCT_{\text{A insert}} / JCT_A$ is to 1, the better.
    \item Assuming task A runs continuously, and task B is insert to run. At this time, the JCT of task B is $JCT_{\text{A insert}}$. The closer the value of $JCT_{\text{B insert}} / JCT_B$ is to 1, the better;
\end{enumerate}
Using an idealized two tasks sharing case study, we demonstrated how \textbf{FIKIT} works compared to Nvidia GPU sharing and exclusive scheduling. Figure ~\ref{fig:caseStudy} shows the schematic diagram of concurrent GPU tasks execution.

\textbf{Nvidia Exclusive Execution Scheme}:
In Nvidia GPU's exclusive mode, it is impossible to concurrently launch multiple tasks. Tasks A and B need to be orchestrated and executed sequentially by an external program. Assume $JCT_A$ and $JCT_B$ are obtained through measurement.

As shown in A,B Exclusive 1 in Figure~\ref{fig:caseStudy}, the external program schedule Task A before task B, $JCT_{\text{A Actual}} = JCT_A$, $JCT_{\text{B Actual}} = \text{Task B waiting time} + JCT_B = JCT_A + JCT_B$.

As shown in A,B Exclusive 2 in Figure~\ref{fig:caseStudy}, the external program schedule Task B before task A, $JCT_{\text{A Actual}} = \text{Task A waiting time} + JCT_A = JCT_B + JCT_A, JCT_{\text{B Actual}} = JCT_B$;

From the data above, we can see that in exclusive mode, the "$JCT_{\text{A Actual}} / JCT_A$" ratio for high-priority task A is not necessarily good, it depends on the order of task execution. If N tasks A happen to execute first, the $JCT_{\text{B Actual}} / JCT_B$ for low priority task B is also hard to predict or even unable to execute (starvation). Therefore, the Nvidia GPU Exclusive execution scheme cannot achieve the concurrent execution goal in evaluation metric 1), performs poorly on metric 2) because it is "unstable", and is also "poor" on metric 3): It is not an ideal concurrent scheme.

\textbf{Nvidia default sharing scheme}:
With Nvidia GPU's default sharing mode for concurrent execution of tasks A and B, the GPU does not know the priorities of A and B. Through testing we find that for different types of tasks A and B, the GPU shared mode execution could be A, B Sharing 1 or A, B Sharing 2 mode as shown in Figure~\ref{fig:caseStudy}.

For A,B Sharing1, $JCT_{\text{A Actual}}$ is significantly larger than $JCT_A$, $JCT_{\text{B Actual}}$ is slightly larger than $JCT_B$. For A,B Sharing 2, $JCT_{\text{A Actual}}$ is slightly larger than $JCT_A$, $JCT_{\text{B Actual}}$ is significantly larger than $JCT_B$;

The above analysis shows that in shared mode, the prioritized execution of task A is not always guaranteed. The order is random and depends on the combination between tasks. For evaluation indicator 1) The performance of the sharing mode is better than that of the exclusive mode, but it is still not ideal, "medium". When task B runs continuously and task A is inserted to run, shared mode cannot guarantee task A preempts the GPU first. So for evaluation metric 2), shared mode performs "poorly". For evaluation metric 3), the low priority task B usually gets a relatively low $JCT_{\text{B Actual}}$, performing "well".

\textbf{FIKIT Scheme}:
With the FIKIT scheme to run tasks A and B concurrently, we schedule tasks A and B to run concurrently with schedules that interleave low priority task B's kernels in tandem with high priority task A's idling gaps. The FIKIT scheme has the ability of preemptive priority scheduling, so that even in the scenario where task B is running continuously and then inserted to run task A, task A can obtain GPU resources in a timely manner to ensure the high priority JCT meets evaluation metrics 1) and 2). As shown in Figure 3, tasks A and B time-share the GPU, and the "$JCT_{\text{B Actual}} / JCT_B$" of task B is also stable and predictable. Ideally, the "$JCT_{\text{A Actual}} /JCT_A$" of Task A will be close to 1. Therefore, FIKIT is the only scheme that satisfies all three metrics simultaneously.

Our third point of view: In the cloud computing environment, only non-intrusive solutions are truly applicable in production. This non-intrusive solution satisfied the following requirments:
\begin{enumerate}
    \item Do not rely on open-sourced GPU drivers, such as Nouveau, because they have suboptimal performance and not widely adopted.
    \item Cracking closed-source GPU driver programs, such as cracking a certain version of Nvidia drivers to improve efficiency, should not be done. Doing so can lead to instability and uncertainty when upgrading to future versions.
    \item Cannot directly modify the client-side source code running in the cloud server. This is not allowed by customers in cloud computing environments.
\end{enumerate}
Our fourth point of view: Any GPU scheduling system needs to keep its total additional overhead to the GPU task execution time acceptably low level, say below 5\%. scheduling technique itself incurs too high of an extra GPU consumption, the effect of scheduling optimization will be offset, resulting in no improvement in GPU resource utilization. Therefore, reducing the scheduling technique's overhead on the GPU is a key prerequisite for achieving scheduling optimization. To minimize FIKIT's impact on GPU task execution time, we divide FIKIT system into two stages when serving a GPU task: profiling phase and execution phase. Utilizing the characteristic of cloud service tasks being repeatedly called many times in a cloud computing environment, this effectively limits high GPU performance overhead caused by the profiling operations.

Based on the four viewpoints, we have designed a FIKIT scheduling algorithm, which can utilize the inter-kernel GPU idle time of the high-priority task to perform low-priority tasks GPU kernel, to fill the wasted computation time of the GPU to maximize the use of computation resources. Through the FIKIT scheduling mechanism, high-priority tasks and low-priority time-sharing can be run on a single GPU at the same time. It prioritizes the execution of high-priority tasks, reduces the JCT of time-sensitive tasks, and performs low-priority tasks during kernel-level GPU idle. Multiple tasks concurrently share the GPU.

\section{System Design}
\subsection{Assumptions and Limitations}
The design of FIKIT was based on the following assumptions and limitations.
\begin{enumerate}
    \item The approach proposed in this paper is mainly applicable to tasks with a huge number of repeated calls in various computing environments. This type of task is common in cloud computing environments and may also appear in enterprise clusters or multi-machine environments. If the number of repeated calls of tasks in this environment is not high, the benefits of the method in this paper will be limited. In this paper, the scope suitable for using the FIKIT scheme is referred to as "multi-tasking and containerized commercial cloud computing environments" or "cloud computing environments".
    \item It is applicable to scenarios where GPU tasks have different priorities. If there are GPU tasks with the same priority, they will share the GPU in a time-slice rotation manner. Scheduling between equal priority tasks is outside the scope of this paper.
    \item The method proposed in this paper is based on GPU time-division multiplexing technology. Therefore, theoretically it cannot be used together with other schemes adopting time-division multiplexing. For example: the Nvidia vGPU/MPS platform cannot use the method proposed in this paper. In addition, many research papers on GPU sharing methods using time-division(Gemini\cite{13}) multiplexing also cannot coexist with the method in this paper.
    \item Theoretically, FIKIT supports GPU hardware from different manufacturers, but the current project implementation and test program only supports Nvidia GPUs and its CUDA API. Theoretically, it should support Pytorch/TensorFlow/CUDA programs, but it has only been tested and verified on Pytorch and CUDA programs, without testing and verification on TensorFlow.
\end{enumerate}
\subsection{Overall design and implementation}
FIKIT achieves priority-based preemption through multiple priority queues, executing low-priority task can immediately pause and give way GPU resources to later coming high-priority tasks.

While high priority tasks are running, FIKIT predicts the high-priority task's Inter-kernel idle time through the kernel statistics recorded in the profiling phase, and schedules multiple low-priority GPU kernels that satisfy the algorithm's requirements for execution during that idle gap, to maximize filling the GPU's computation time.

\begin{figure}[tp]
    \centering
    \includegraphics[width=\columnwidth]{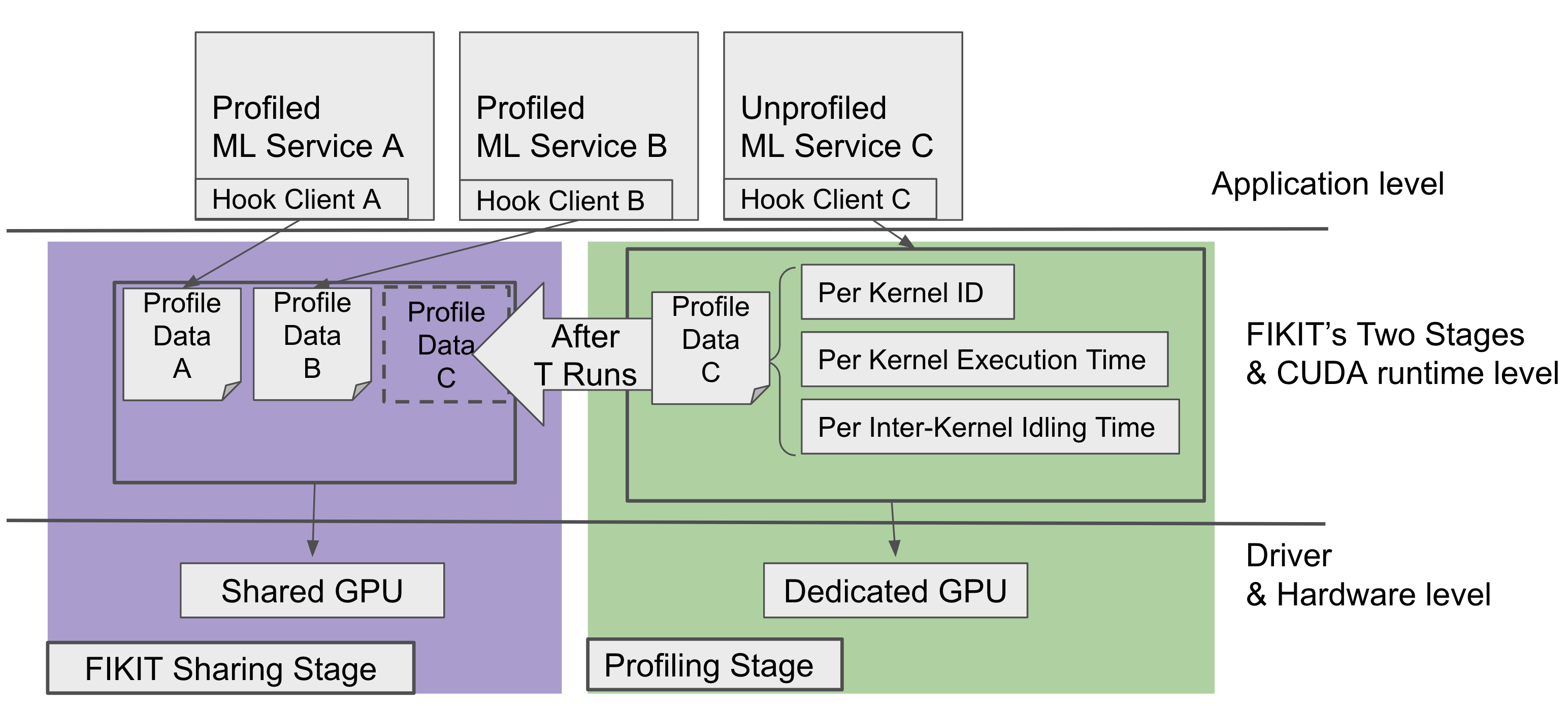}
    \caption{A service's execution is divided into two phases: measurement and FIKIT execution phase.}
    \label{fig:overallDesign}
\end{figure}

The implementation of the FIKIT framework is divided into three parts:
\begin{enumerate}
    \item Recompiled PyTorch/TensorFlow library, it is based on the standard PyTorch/TensorFlow library, without modifying any of the original code, only by adding the compilation option "-rdynamic" to the original library recompiled.
    \item Hook client, which is a preload library written in cuda c/++ language. Each task is loaded in the form of Preload library at the start of the process. CUDA hook technology is used to intercept each GPU kernel executed in the process. kernel id can be resolved through the "recompiled PyTorch/TensorFlow library".
    \item FIKIT scheduler, an independent process written in C/C++. With task processes and Hook clients can be deployed on different machines, the hook client communicates with the FIKIT Scheduler through UDP messages.
\end{enumerate}

The hook client is responsible for intercepting kernel information and sending it to the scheduler. After the scheduler receives the kernel information, it calculates the scheduling strategy and sends the scheduling instructions back to the Hook client. The Hook client chooses when to send it to the GPU to execute the kernel according to the instructions of the scheduler. This design realizes the interception and scheduling control of GPU kernel through Hook technology and adopts a distributed client-server model. The client is responsible for kernel interception and the server is responsible for kernel-level scheduling. The two exchange and coordinate information through UDP communication.

Based on the characteristics of the cloud computing environment to run the same GPU task repeatedly, the program is divided into two phases: measurement and FIKIT execution phase. As shown in the Figure~\ref{fig:overallDesign}.

Task A and Task B (tasks with profiled kernel-level statistics loaded in scheduler) use the FIKIT scheduling algorithm, which can be executed concurrently on the same GPU.

The new task C (a task without measurement data) first executes in measurement phase. After T measurements ($T\in[10,1000]$), the measurement data is generated using the statistical method. Task C runs again with the measurement data will enter the FIKIT scheduling execution phase. Measuring operations has a large impact on the task execution time. This design eliminates the impact of measurement on task efficiency by dividing into measurement and FIKIT execution stages. This is because the measurement stage only occupies a very small portion of the total execution time, which can be proven by the quantitative analysis below.

Let task A be executed a total of N times, where $N_m$ times are executed in measurement mode and $N_f$ times are executed in FIKIT mode.
\begin{itemize}
    \item $JCT_m$, the average completion time of task A under the measurement phase.
    \item $JCT_f$, the average completion time of task A during the FIKIT phase.
    \item $JCT_{overhead}$ defines the completion time overhead ratio due to the measurement phase.
    \item $JCT_{overhead}= JCT_m / JCT_f$, test results show that $JCT_{overhead}$ is between 1.3 \textasciitilde 1.7, i.e. $max(JCT_m / JCT_f) = 1.7$.
    \item $JCT_{avg}$ The average JCT for N times in the measurement phase and FIKIT phase.
\end{itemize}
From the definition above, we can conclude that
\begin{align*}
    N&=N_m+N_f \\
    JCT_{avg}&=(N_m * JCT_m+N_f * JCT_f)/(N_m+N_f)\\
\end{align*}
After simplification, this gives
\begin{align*}
    JCT_{avg}=JCT_f+0.7\ast(N_m/N)\ast JCT_f 
\end{align*}
When $N\gg N_m$
\begin{align*}
    JCT_{avg}\simeq JCT_f
\end{align*}
\textbf{Identifying each kernel}
One of the key techniques to implement FIKIT scheduling algorithm is to identify each kernel using "kernel id". No previous work has addressed kernel identification problem. Without mechanisms to precisely identify kernels, some studies \cite{13} have chosen online realtime measurement of kernel execution time in their implementation of kernel-level GPU schedulers. Online real-time measurement will greatly increase the performance overhead during GPU kernel execution, which will lower the performance optimization brought by the scheduler, and may even worsen the overall performance. This work implements an efficient method to identify GPU kernel functions at runtime, which allows us to move kernel measuring offline, and then apply offline measurement statistics during runtime. This greatly improves the scheduling efficiency of the scheduler at runtime.

\begin{figure}[tp]
    \centering
    \includegraphics[width=\columnwidth]{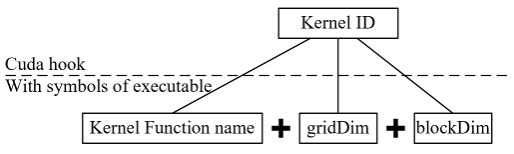}
    \caption{The construction of a Kernel ID.}
    \label{fig:kernelID}
\end{figure}

 A kernel's function name, dimension of each thread block within the grid and the allocated dimension of grid makes up of its kernel ID. As shown in the Figure~\ref{fig:kernelID}.
There are block and gid dimensions available through the intercepted CUDA runtime API call. However, no cuda API exposes kernel function names (for the Pytorch/Tensorflow default release version). After some research, recompiling the version of Pytorch/Tensorflow by adding the -rdynamic compile option to Pytorch/Tensorflow solves this problem. Based on this recompiled version, the cuda hook technique can get the kernel name, thus solving the problem of generating the Kernel ID. the profiler requires the service is loaded with a dynamic symbol exported The profiler requires the service is loaded with a dynamic symbol exported compilation of the ML framework i.e. PyTorch or TensorFlow, which can be loaded by cloud providers, making it available for all hosted services using the framework. The profiler can reconstruct kernel function names through reading symbolised backtrace.

With performance evaluation, service using the dynamic symbol exported version of PyTorch, and Tensorflow library incurs less than 2\% overhead which can be safely ignored. (This needs to be changed to show that there is almost no difference in performance between them, and see the test results in \ref{sec:rdynamicvsbase}.)

\begin{figure}[tp]
    \centering
    \includegraphics[width=\columnwidth]{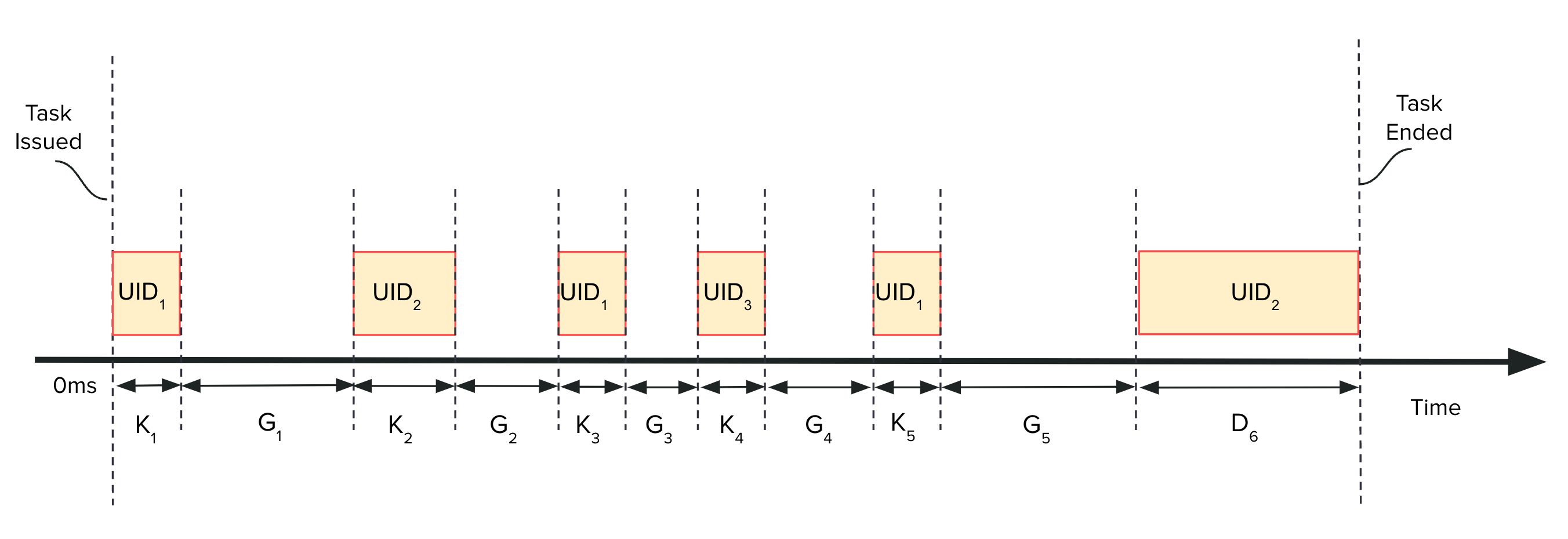}
    \caption{There can be kernels of different execution duration but with the same kernel ID.}
    \label{fig:differentKernel}
\end{figure}

The advantage of using kernel is that it decouples the execution order of the kernel to its measured statistics. But there is one shortcoming of the current kernel ID design. Although the kernel ID is effective in identifying kernels with their specific kernel functions and parallelization level (block and grid scale), which effectively identifies kernels by their computation intensities. The kernel ID does not record a kernel function's input information. The kernel function inputs are of 'void' type at the CUDA runtime level, making it hard to reconstruct without inspecting the service source code. In order to work in a cloud environment without touching source code of hosted ML services, FIKIT tradeoff some kernel identification precision to the generality of this solution. There can be kernels of different execution duration but with the same kernel ID because the current kernel ID cannot identify different input scales of kernels calling the same kernel function with the same parallelization level. As illustrated in the figure~\ref{fig:differentKernel}, the first and second times of executing kernel K2 have different duration D1 and D2. To mitigate this shortcoming, the mapping uses the average of all durations associated with this kernel ID across T runs at the profiling stage and at the FIKIT stage the scheduling algorithm has a dynamic duration and idling prediction mechanism that will be explained in later sections.

\textbf{Measuring the execution and idle time of kernel}
Profiled data includes the Kernel id, Kernel execute time, and Inter-kernel idle time.
To measure the Profiled data about kernel, key challenges of the profiler are: 
\begin{enumerate}
    \item CUDA driver does not provide kernel profiling API and the profiler does not have access to the service source code, the profiler needs generate a custom ID to Identify kernels. 
    \item Programs running on CPU launches kernels asynchronously to GPU, the profiling mechanism need to measure the actual execution start and end of a kernel on GPU.
\end{enumerate}

To obtain the execution time of kernels, REEF\cite{30} is implemented based on AMD ROCm \cite{42} (an open-source GPU computing platform), and requires modifying the binary or assembly code of the GPU kernel in order to measure the kernel execution time. It only supports the AMD Radeon Instinct MI50 GPU device, and lacks generality. There are also some studies(TimeGrapsh\cite{10}) trying to measure the kernel by modifying open-sourced drivers (such as nouveau\cite{9}), but open sourced drivers are  hardly used in the cloud computing environment and thus methodologies developed from these drivers have large gaps from adaptations. Using CUDA API hook\cite{3} technique, FIKIT scheduling algorithm preloads custom profilers with CUDA runtime API\cite{3} invocations to measure kernels' execution. Without depending on open-sourced GPU drivers, this method can measure kernels' execution precisely. Of course, using this method will generate large performance overhead in overall task JCT. Through our tests, we found that in most cases, it will slow down the job completion time of a GPU task by 20\%-80\%. To reduce the overhead of measuring kernel execution time, Gemini\cite{13} adopts methods such as "incidentally" inserting monitoring events when the application synchronize to reduce the performance overhead caused by hooks. The effectiveness of this method depends on the number of synchronization events in a task and thus cannot guarantee stable performance and has weak generality.

The method proposed in this paper is also based on CUDA API hook technique, however, it utilizes the characteristics of the cloud computing environment. With GPU kernel identification technique, the new method enables offline measurement of GPU kernel function execution time to be applied in an online real-time manner. By separating the measurement operations from the FIKIT scheduling process, FIKIT eliminates the impact of measuring kernel time from the scheduling algorithm, thereby greatly improving the efficiency of FIKIT scheduling.

With the kernel measuring method of this work, GPU tasks lose 20\%-70\% of JCT performance during the measurement stage. At the real-time sharing stage, GPU tasks, however, FIKIT only incurs JCT overhead within 0.1\%-5\%,and less than 3\% in most scenarios.

The reasoning of the effectiveness of the proposed kernel measuring method is as follows:

When running GPU tasks in a cloud computing environment, the same cloud service program is usually called repeatedly, and the number of repeated invocations can be up to 100,000 times or more. According to this characteristic of cloud computing, the FIKIT program framework is divided into two phases: running service with profiling phase and running service with FIKIT scheduling phase. The purpose is to perform measuring operations with large GPU overheads in a limited number of task executions, and to make efficient GPU scheduling in the following large numbers of repeating service invocations under FIKIT scheduling phase. When a new GPU task (a task without measurement data) is waiting to execute, we will measure it. During the measurement phase, the GPU task occupies the GPU exclusively. After N (usually N within 10-1000) execution with measurements, the profiled kernel-level data of the program is statistically processed and loaded into memory. When this GPU task is waiting to be executed again, since it has existing profiled data, no more measuring operations will be performed on it. The FIKIT scheduling policy will execute it con- currently according to its priority, and its performance will be close to a normal invocation afterwards. The GPU task will usually lose 20\%-80\% performance in JCT during the measurement phase, but compared to 100,000+ normal invocations, the 20\%-80\% performance loss in less than 1000 invocations can be ignored.

\textbf{Data acquisition and statistical output during the measurement phase}: 
\begin{figure}[tp]
    \centering
    \includegraphics[width=\columnwidth]{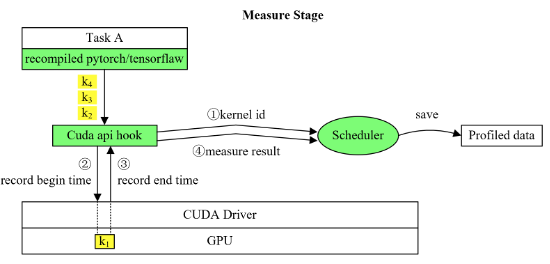}
    \caption{Kernel-level Measurement}
    \label{fig:measureStage}
\end{figure}
As shown in the Figure~\ref{fig:measureStage}, in the measurement stage, the task exclusively uses the GPU, intercepts each kernel call through cuda hook technology, obtains the kernel name and related kernel parameters through recompiled pytorch/tensorflow, and generates kernel id information. In addition, cuda event is used to record the start and end times of kernel execution and calculate the execution time of the kernel and the time from the end of the kernel to the start of the next kernel, that is, inter-kernel idle time. 

Profiler measures records per kernel measurement for T runs and outputs a statistical result.

The kernel-level data that needs to be collected through measurements includes:
\begin{enumerate}
    \item Kernel ID for each executed kernel in T runs, denoted as $ID_{t,i},0<t\le T,0<i\le N_t$. $N_t$ is the total number of kernels in t-th ML task run on the GPU. The Kernel ID is independent of kernel's sequential index within a task. For example, in t-th run, the first and fifth kernel's ID, $ID_{t,1}$ and $ID_{t,5}$ can be identical, indicating they are repeating kernel execution. The Kernel ID is also independent of index of t-th run. For example, the first run and second run's kernels can have identical kernel ID. $ID_{1,i}=ID_{2,j}$ exists,$0<i\le N_1,1<j\le N_2$.  
    \item Per kernel execution time, denoted as $K_{ID_{t.i}},0<t\le T,0<i\le N_t$. From a ML task's start to end, the profiler will record for example: $K_{ID_{t,1}}=2ms,K_{ID_{t,2}}=3ms,\ldots K_{ID_{t,N_t}}=10ms$.
    \item GPU idling time, $G_{ID_{t,i}},0<t\le T,0<i<N_t$. It is the duration from each kernel ends to the next kernel starts. For example, after $K_{ID_{t,1}}$ ends and before $K_{ID_{t,2}}$ starts, the time is recorded as idling time $G_{ID_{t,1}}$. From a t-th run of the ML task's start to end, there is $N_t-1$ idling time. The profiler will record for example: $G_{ID_{t,1}}=10ms,G_{ID_{t,2}}=5ms,\ldots G_{ID_{t,N_t-1}}=8ms$.    
\end{enumerate}

Based on the measurement data collected, we calculate and output the following statistical results:
\begin{enumerate}
    \item 	The set of unique kernel IDs denoted as $S_{UID}$. The ID sequence defined above account complete kernel execution record. However, as a ML task would call the same type of kernel within and across each run. It is not a set that is readily available for FIKIT scheduler's prediction. $S_{UID}$ uniquify sequence of executed kernel from the ID sequence and form such a set of Kernel IDs.
    \item 	the average execution time of kernels with the same ID across $T$ runs, denoted as $SK$. \\
        $SK=\left[{SK}_j\right],\forall j\in S_{UID}.\left|SK\right|=\left|S_{UID}\right|$.\\
        $SK_j=\frac{\sum_{0<t\le T}\sum_{0<i\le N_t} K_{ID_{t,i}}\ast\delta (ID_{t_i},j)}{\sum_{0<t\le T}\sum_{0<i\le N_t}1\ast\delta (ID_{t_i},j)}$ \\
        $\delta$ is the Kronecker delta function, which equals 1 if the two arguments are equal, and 0 otherwise.\\
        For example, a ML task is measured 2 times. For a given kernel ID, $j$. In the first run, the first and fifth kernels have $ID_{1,1}=ID_{1,5}=j$. In the second run the second and sixth kernels have $ID_{2,1}=ID_{2,5}=j$. $SK_j=\left(K_{ID_{1,1}}+K_{ID_{1,5}}+K_{ID_{2,2}}+K_{ID_{2,5}}\right)/4$ is the average kernel execution time of the kernel with same ID repeated across 2 runs.
    \item the average GPU Idling time after the kernels with the same ID across $T$ runs, denoted as $SG$.\\
        $SG=\left[{SG}_j\right],\forall j \in S_{UID}. \left|SG\right|=\left|S_{UID}\right|$.\\
        $SG_j=\frac{\sum_{0<t\le T}\sum_{0<i\le N_t-1} G_{ID_{t,i}}\ast\delta (ID_{t_i},j)}{\sum_{0<t\le T}\sum_{0<i\le N_t-1}1\ast\delta (ID_{t_i},j)}$ \\
        $\delta$ is the Kronecker delta function, which equals 1 if the two arguments are equal, and 0 otherwise.
        For example, a ML task is measured 2 times. For a given kernel ID, $j$. In the first run, the first and fifth kernels have $ID_{1,1}=ID_{1,5}=j$. In the second run the second and sixth kernels have $ID_{2,1}=ID_{2,5}=j$. $SG_j=(G_{ID_{1,1}}+G_{ID_{1,5}}+G_{ID_{2,2}}+G_{ID_{2,5}})/4$ is the average GPU idling time after executing these same type of kernels repeated across 2 runs.
\end{enumerate}

During the measurement stage, the Profiling data of each task is generated by the following methods:
1. According to the process name and startup parameters of the task, the Task Key is generated as the unique identifier of the task.
2. Obtain the unique ID($SK_{UID}$) and scheduling group ID($SG_{UID}$) of each Kernel when the task is executed by measuring.
3. Associate Task Key with the measured $SK_{UID}$ and $SG_{UID}$.
4. Finally, Task Key is used as the keyword to record the Profiling data of all Kernel execution information ($SK_{UID}$, $SG_{UID}$, etc.)in the task. The outputted profiled result of a service is  $Task Key= (SK, SG)$.

In summary, the measurement stage generates a unique Key for each task, through which Kernel information data is associated to form task-level Profiling data. FIKIT scheduling algorithm will use the Profiling.

\textbf{FIKIT scheduling design}: The core idea of FIKIT is to take advantage of the idle time when the high-priority task GPU Kernel is running to execute the low-priority task GPU kernel.
FIKIT stage has the following core technical design:
\begin{enumerate}
    \item Loading profiled data
    \item Executing kernels by task-level priorities
    \item Fill inter-kernel idling time 
    \item Real time feedback to stop error propagation
    \item Task Switching
\end{enumerate}
At the FIKIT stage, there is one central controller scheduling each ML service processr's kernel execution through a client hooked with CUDA runtime library.  

\textbf{Loading profiling data}: Task Key, as a unique identification of the task. From the measurement output data, filter out the profiling data matching the Task Key, which will be used in the FIKIT scheduling algorithm.
\textbf{Executing kernels by task-level priorities}: 
\begin{figure}[tp]
    \centering
    \includegraphics[width=\columnwidth]{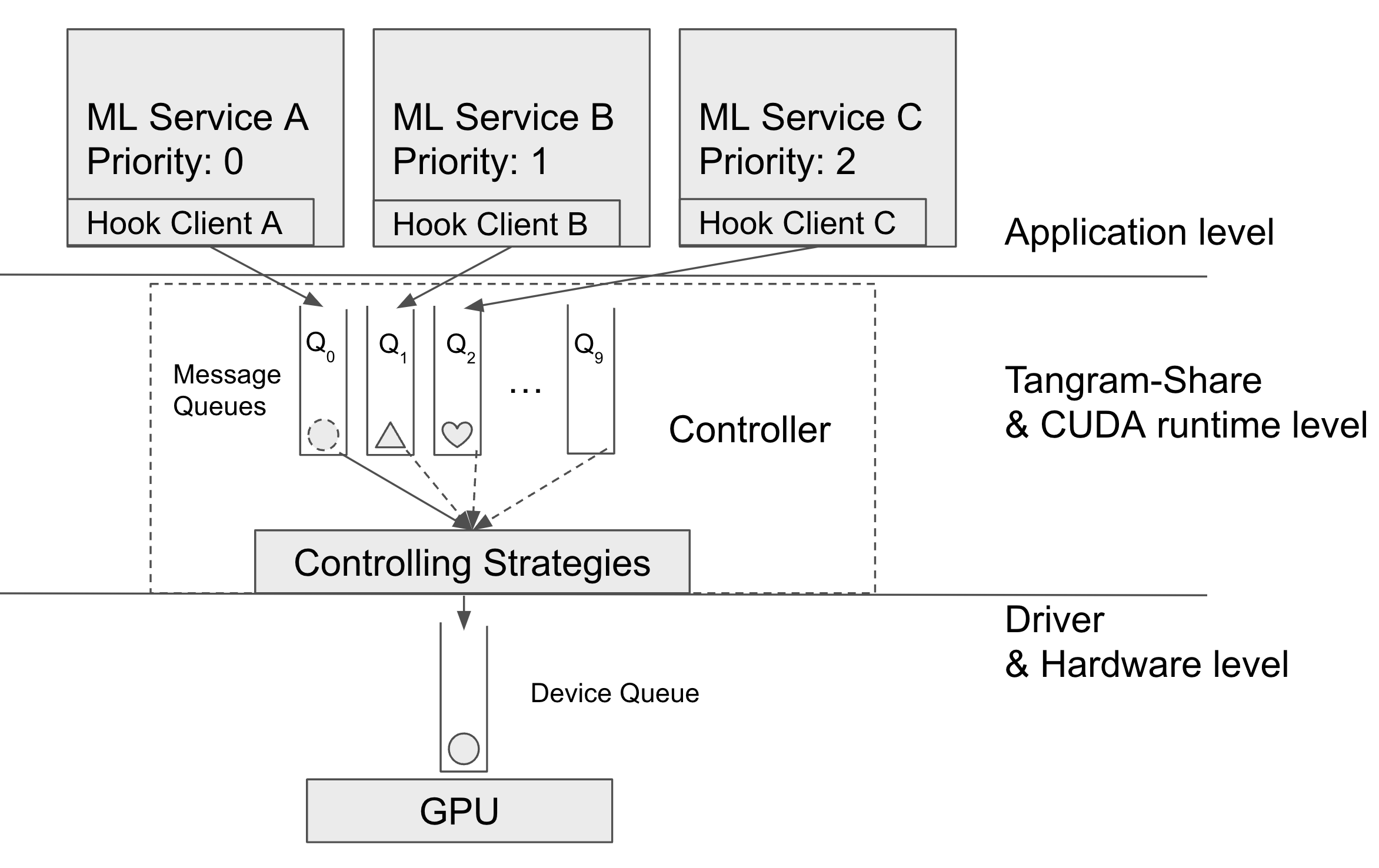}
    \caption{Priority Queues}
    \label{fig:priorityQueues}
\end{figure}
As shown in Figure~\ref{fig:priorityQueues}, the scheduler supports up to 10 priority queues (labeled Q0-Q9). It implements priority scheduling by scanning queues in order of priority from highest to lowest (Q0 to Q9).
To be specific:
\begin{enumerate}
    \item The queue with the highest priority is $Q_0$,and the queue with the lowest priority is $Q_9$.
    \item The scheduler scans the queues in the order of $Q_0 \rightarrow Q_1 \rightarrow ... \rightarrow Q_9$. If there are tasks waiting to be scheduled in a high priority queue, tasks will be scheduled with priority from the high priority queue. Only when the high priority queue is empty will tasks from lower priority queues be considered for scheduling.
\end{enumerate}
Therefore, the algorithm can ensure that high-priority tasks will be scheduled first.

The workflow of the scheduling mechanism is shown in Figure~\ref{fig:priorityQueues}:
\begin{enumerate}
    \item A task need to be started with one of 0-9 priority.
    \item Intercept the kernel call of the task through the hook client and push the kernel to the corresponding priority queue according to the priority specified by the task.
    \item The scheduler scans the queues in order of priority from highest to lowest (0-9).
    \item Select a kernel from the non-empty highest priority queue and place it in the GPU execution queue.
    \item The GPU executes the kernel in the same queue in FIFO order.
\end{enumerate}

The priority-based GPU scheduling is completed by knowing the kernel call and priority information through hook, and the scheduler decides the execution order of the kernel according to the priority, and the GPU executes the kernel in the final order.

\textbf{Fill inter-kernel idling time process}: 
\begin{figure}[tp]
    \centering
    \includegraphics[width=0.7\columnwidth]{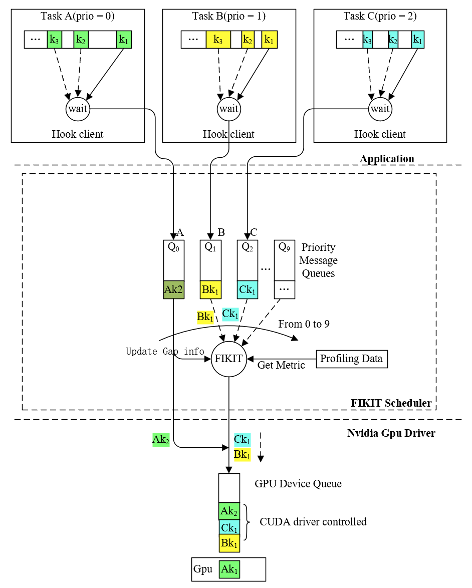}
    \caption{Process of filling inter-kernel idling time}
    \label{fig:fillIdling}
\end{figure}

As shown in the Figure~\ref{fig:fillIdling}, A, B and C are three tasks with priority from high to low. When they are executed concurrently, each task will be intercepted by its own hook client when the kernel calls, and the kernel task will be placed in the waiting state. The highest priority task A holds the GPU, and its first kernel, $Ak_1$, goes directly to the GPU execution queue. The second highest priority task B cannot obtain the GPU, and its kernel $Bk_1$ is put into priority queue $Q_1$. The lowest-priority task C cannot obtain the GPU, and kernel $Ck_1$ is put into priority queue $Q_2$. When a kernel is added to any priority queue, the scheduler triggers a priority scan. From high-priority $Q_0$ to low-priority $Q_9$, the kernel satisfying BestPrioFit algorithm is placed into the GPU queue. When the idle time from $Ak_1$ to $Ak_2$ is long, the scheduler does not wait for $Ak_2$ of task A, but inserts low-priority $Bk_1$ and $Ck_1$ into the GPU queue. Therefore, their execution sequence is $Ak_1-Bk_1-Ck_1-Ak_2$, which realizes the function of "Fill inter-kernel idling time" (FIKIT).

The algorithm process of FIKIT is shown in Algorithm 1.
\begin{figure}[tp]
    \centering
    \includegraphics[width=0.9\columnwidth]{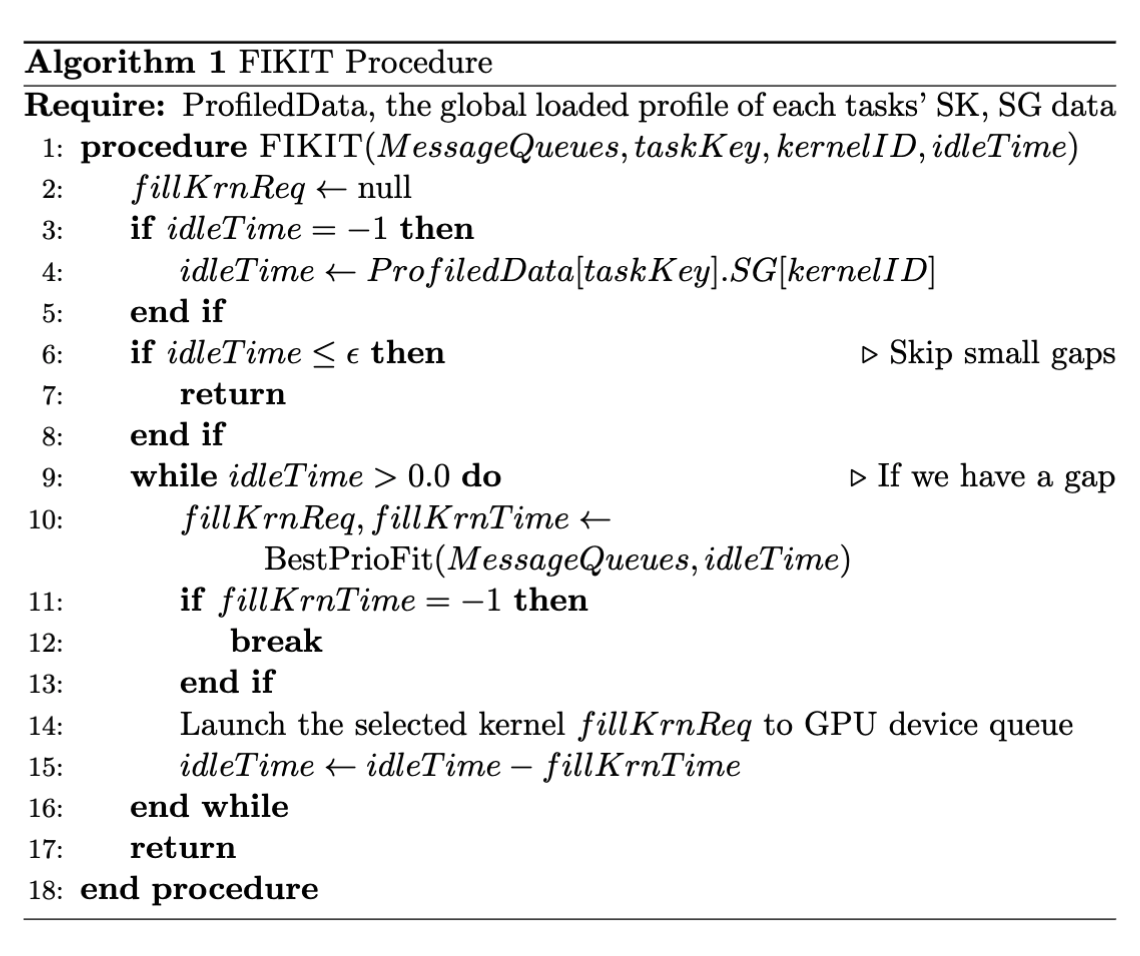}
    \caption{Algorithm 1: FIKIT Procedure}
    \label{fig:algo1}
\end{figure}
\begin{figure}[tp]
    \centering
    \includegraphics[width=0.9\columnwidth]{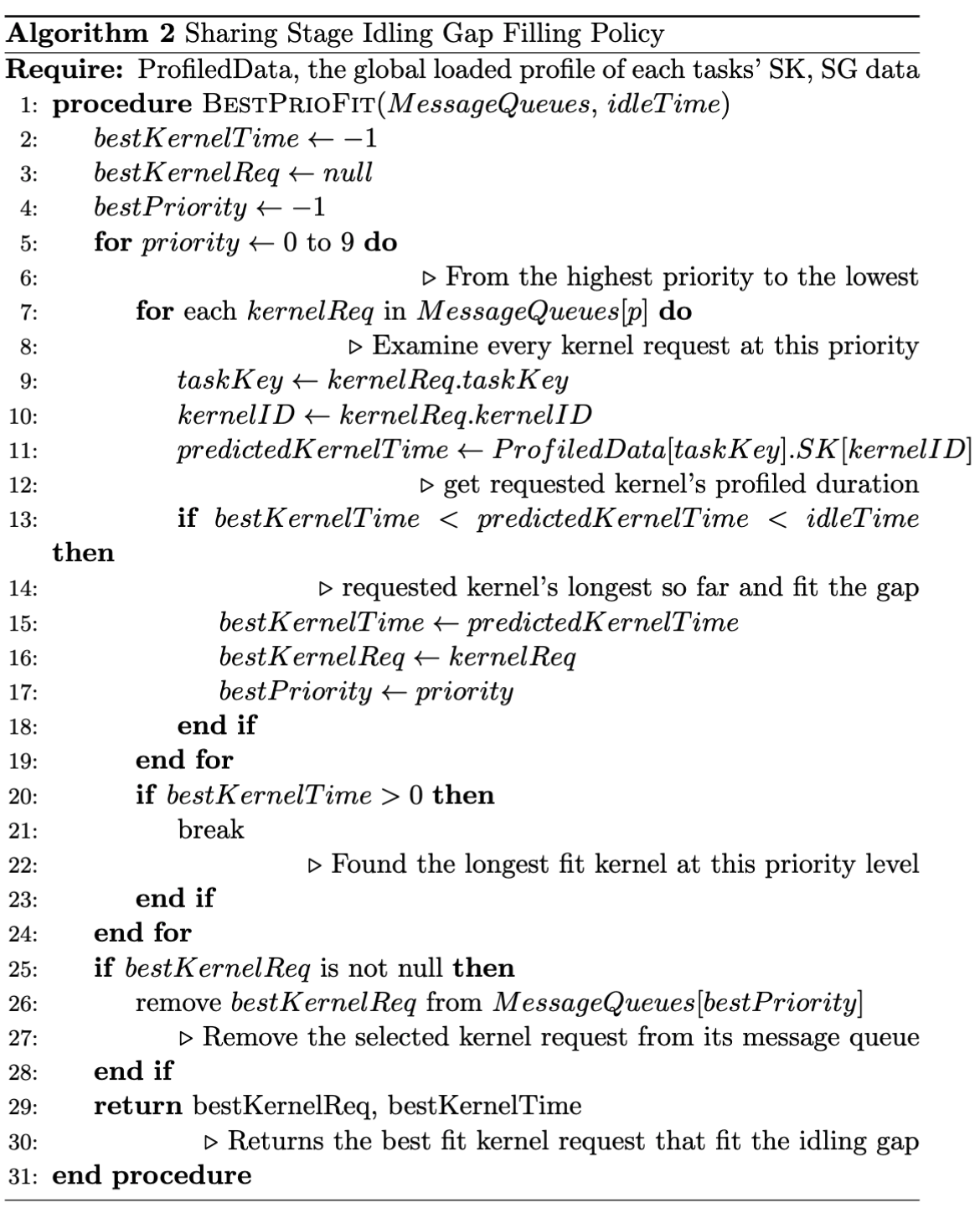}
    \caption{Algorithm 2: Sharing Stage Idling Gap Filling Policy}
    \label{fig:algo2}
\end{figure}
When the task's kernel running on the GPU has an idle time after its execution, the \textit{FIKIT} procedure will be invoked. It repeatedly calls the \textit{BestPrioFit} procedure to find and allocate the most suitable kernel requests (from the highest priority to the lowest) to fill this gap. The \textit{FIKIT} procedure continues until either the idling time is fully occupied or there are no more suitable kernel requests. Essentially, \textit{FITKIT} aims to maximize GPU utilization by filling its idle times with as many low-priority kernels as possible.

In the \textit{FITKIT} procedure (Algorithm 1, fig~\ref{fig:algo1}), it focuses on efficiently utilize the idle time (gap) that occurs when a high priority kernel is running on the GPU. If the running kernel's idle duration hasn't been looked up yet, the procedure read the profiled data associated with the task to get the idle time (line 3-5). A kernel launched on the GPU typically costs 0.1ms to 2ms. The function avoids filling negligible idle gaps that is smaller than 0.1ms by skipping them (line 6-8). The procedure repetitively calls the \textit{BestPrioFit} procedure, each time aiming to find the most optimal kernel request that can fill part or the entirety of the gap (line 9-16). For every suitable kernel found, they are launched to GPU device queue for execution (line 14), and the idle duration is revised to reflect the newly occupied time (line 15). This mechanism ensures that the idle times of the GPU are efficiently populated by running as many low-priority kernels as possible, thus optimizing GPU utilization.

The \text{BestPrioFit} procedure (Algorithm 2, fig~\ref{fig:algo2}), on the other hand, encapsulates the logic for selecting the most suitable kernel to fill a given idle duration. The intuition behind \textit{BestPrioFit} is to find a kernel that can best fit the current idle gap from a list of waiting kernels organized by priority levels. The term "best fit" here implies finding a kernel with 1) the highest priority that does not exceed the idling time 2) whose execution time is closest to the remaining idle time among candidates of the same priority. By iterating through the priority levels from highest to lowest and examining each kernel request, \textit{BestPrioFit} ensures that higher-priority kernels are given preference over lower-priority ones when filling the idle gap.

The process in \textit{BestPrioFit} begins by iterating through the priority levels (line 5), examining each kernel request at the particular priority level (line 7) to find a kernel whose execution duration is the longest so far and fits within the remaining idle gap (lines 13-18). If a suitable kernel is found, it is dequeued from the waiting queue, and the procedure returns the best fit kernel request along with its duration (lines 25-29 in Algorithm 2).

By synergizing the \textit{FITKIT} and \textit{BestPrioFit} procedures, the algorithm aims to strategically fill the GPU holding task's idling gaps with lower-priority kernels, thus striving to maintain a high level of GPU resource utilization while respecting the priority order of task execution.

\begin{figure}[tp]
    \centering
    \includegraphics[width=\columnwidth]{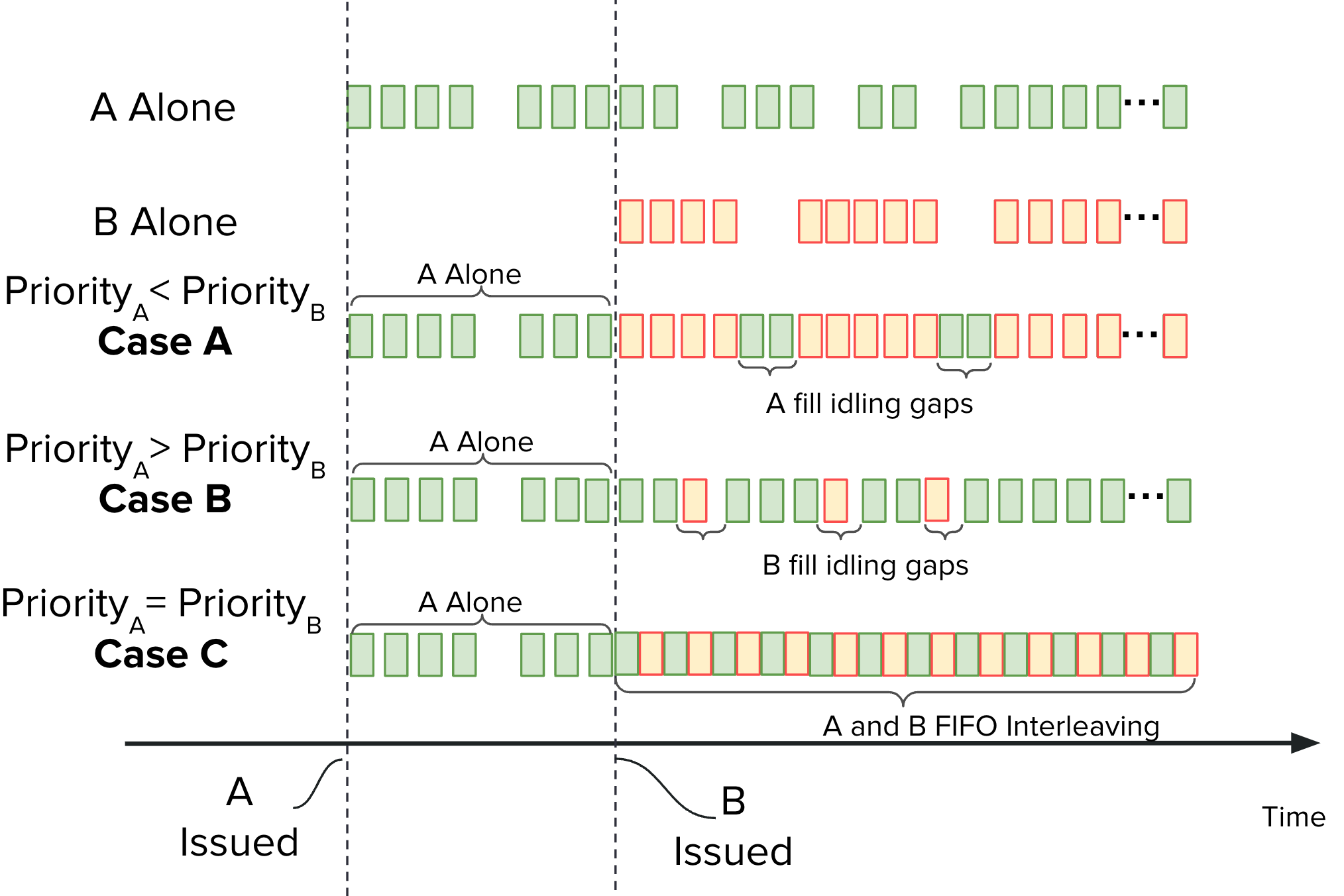}
    \caption{A Case study of Idling Gaps filling with Priority}
    \label{fig:fillGapwPriority}
\end{figure}

When tasks are not started at the same time, the scheduler and FIKIT procedure can work without problem. As illustrated by figure~\ref{fig:fillGapwPriority}, we use a two-tasks sharing example, explaining how scheduler with FITKIT algorithm implements priority-based GPU sharing. Task A starts first, and task B is issued later. There are three possible cases: 1) A has lower priority than B (figure~\ref{fig:fillGapwPriority}, case A), 2) A has higher priority than B (figure~\ref{fig:fillGapwPriority}, case B), and 3) A and B have the same priority (figure~\ref{fig:fillGapwPriority}, case C).

In Case A, the scheduler withholds the next launching kernel of A to let the later arriving high-priority B's kernel run on the GPU instead. Task A's remaining kernels are then executed during the GPU idling time between B's kernels using FIKIT procedure. Without this type of task scheduling, the high priority task B would suffer a high JCT overhead as it would wait for previously launched low priority task A to complete before it can start using the GPU, violating the priority SLO. However, the low to high priority task switching solves the priority inversion problem.

In Case B, the the scheduler ensure high priority task A kernels are scheduled first. When there is a GPU idling time between A's kernels, the scheduler fills it with low priority task B's kernel.

The last type C is when tasks sharing the GPU have equal priority. It shares the device among tasks like the original CUDA GPU sharing, the scheduler will dequeue task A and B's kernels non-deterministically. Their kernels are interleaving based on their launching FIFO order.

From the example above, we demonstrate that regardless of tasks' issuing orders FIKIT ensure a priority-based concurrent tasks schedule and utilizes wasted GPU idling time within a single task. 

\begin{figure}[tp]
    \centering
    \includegraphics[width=\columnwidth]{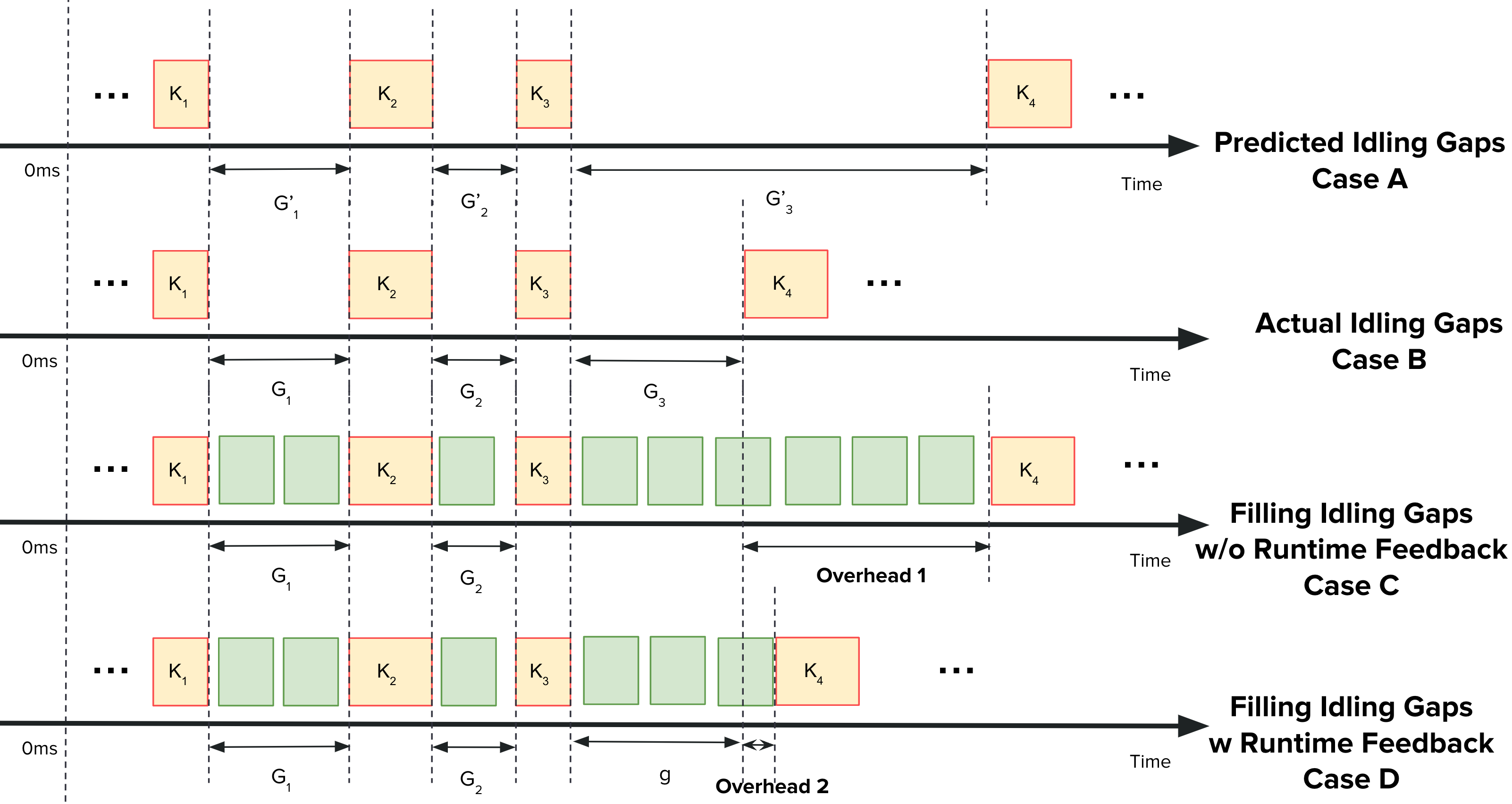}
    \caption{Runtime Feedback}
    \label{fig:runtimeFeedback}
\end{figure}
\textbf{Real time feedback to stop error propagation}: When the controller is using profiling stage results to infer runtime actual kernel and idle duration, the errors of one idle gap prediction can propagate. As illustrated in Figure~\ref{fig:runtimeFeedback}, when a predicted idling time $G'_3$ (shown in case A) overestimates the actual GPU gap, $G_3$ (shown in case B), between two high priority kernels, the controller would schedule more kernels than this idling time $G_3$ can fit, slowing down high priority jobs by overhead 1 as shown in case C. As more and more gap filling procedures occur, the prediction and scheduling error will aggregate linearly. The controller will eventually schedule low priority kernels out of sync with the actual task's idling gap. 

In order to utilize the GPU filling mechanism at runtime, one must solve the error propagation problem described above. We solve this with a real-time feedback and error correction mechanism as illustrated by Case D in Figure~\ref{fig:runtimeFeedback}. The controller dynamically adjusts GPU idling time prediction, $g$, based on real time task execution. While the initial prediction is based on the profiling stage's measurement the scheduler knows the actual end of the idling gap when the next kernel launching request arrives. This is the early stopping signal for FITKIT procedure to immediately stop scheduling low priority kernels since the idling gap has finished earlier than the prediction. However, this method cannot remove the overhead completely. Idling time filling kernels that has been scheduled to the GPU device queue before the early stopping event happen cannot be removed. For example, as illustrated by Figure~\ref{fig:runtimeFeedback}, an idling gap is predicted to be $G'_3$ long by profile stage measurements. The scheduler can fill 6 low priority kernels to fill the idling time. However, the next high priority kernel, $K_4$, actually arrived after $G_3$ duration. FIKIT's controller will not schedule more filing kernels after the third kernel since the runtime event updates the remaining idling time to zero. The controller immediately gives the GPU to the high priority kernel. Without the real time feedback, a purely profile-based scheduler would still fill in $G'_3 - G_3$ worth of low priority kernels and result in a long delay of high priority tasks. However, the mechanism cannot avoid overhead from the already scheduled to GPU's third kernel as illustrated as overhead 2. This dynamic early stopping mechanism mitigates the variation between profiled and actual idling time at runtime, result in the overestimation's overhead decrease from overhead 1 to overhead 2.

\section{Evaluation}
\begin{table*}[htb]
    \centering
    \small 
        \begin{tabular}{|l|l|l|}
        \hline
        \textbf{Task Type} & \textbf{Model} & \textbf{Dataset}  \\
        \hline
        Image - Semantic & fcn\_resnet50 \cite{50} & Google Street View \cite{googleStreetView} \\
        & fcn\_resnet101 \cite{50} &  \\
        & maskrcnn\_resnet50\_fpn \cite{51}  & \\
        & deeplabv3\_resnet50 \cite{52}  & \\
        & deeplabv3\_resnet101 \cite{52} &  \\
        & keypointrcnn\_resnet50\_fpn \cite{51}  & \\
        \hline
        Image - Classification & resnet50 \cite{53}  &  Google Street View \cite{googleStreetView} \\
        & resnet101 \cite{53} &  \\
        \hline
        Image - Object Detection & fcos\_resnet50\_fpn \cite{54} & Google Street View \cite{googleStreetView}   \\
        & fasterrcnn\_resnet50\_fpn \cite{55} &   \\
        & Alexnet \cite{56} &   \\
        & vgg16 \cite{57} &  \\
        \hline
        \end{tabular}
        \caption{DNN Models}
        \label{table:models}

\end{table*}

In this section, we design several experiments to validate the efficiency and effectiveness of FIKIT scheduling architecture.

\subsection{Experimental Setup}
\begin{enumerate}[label=(\roman*)]
    \item \textbf{Experimental scheme I, "-rdynamic" vs "base"}

    In FIKIT, we use the pytorch/tensorflow recompiled by the "-rdynamic" compilation switch. Therefore, we need to evaluate its performance compared to the default compiled pytorch/tensorflow.
    \item \textbf{Experimental Scheme II, Single ML Inference Service FIKIT vs Base}
    
    Under FIKIT and NVIDIA default modes, we host the same ML inference service and measure its tasks' JCT. In a single service scenario, there is no tasks sharing and we evaluate the impact of hosting the service under FIKIT mode compared to the base environment. The overhead of FIKIT mode compared to default mode should be less than 5\%. 
    \item \textbf{Experimental Scheme III, Single ML Inference Service FIKIT Sharing Stage vs Measuring Stage}

    Under FIKIT sharing stage and measuring stage, we host the same ML inference service and measure its tasks' JCT. The goal is to evaluate the effectiveness of limiting kernel-level measurement overhead through dividing FIKIT into measuring and sharing stage. 
    \item \textbf{Experimental Scheme IV, Multiple ML Inference Services Sharing}

    Multiple tasks are executed concurrently in a multiple ML Inference services sharing scenario to compare the performance of FIKIT scheduling mode with default GPU sharing and exclusive modes, respectively. Assume that service A perform high priority tasks and service B performs low priority tasks.
    \begin{enumerate}
        \item When service A and service B are running concurrently on the same GPU, how much the JCT of A's tasks will be improved in the FIKIT scheduling mode compared to the GPU sharing mode.
        \item When service A and service B are running concurrently on the same GPU, how much will the JCT of A's high priority task decrease in FIKIT scheduling mode compared to the case of A's tasks occupying the GPU exclusively; how much will the JCT of B's low priority tasks improve in FIKIT scheduling mode.
        \item When service B runs low-priority tasks continuously in the background, A starts high-priority tasks. We evaluate the JCT of A's tasks in the FIKIT mode compared with the GPU sharing mode and the exclusive model.
        \item When service A runs high priority tasks continuously in the background, service B starts the low-priority tasks. We evaluate if B's tasks JCT is stable and predictable. 
    \end{enumerate}

\end{enumerate}

\textbf{Hardware and software environments} 
Our experiment used a server with an AMD Ryzen 7 5700X 8-core CPU, 32GB of RAM, and an NVIDIA GeForce RTX 3090 (24GB of device memory). The machine runs Ubuntu 20.04.5 LTS operating system and uses NVIDIA driver version 525.147.05 (525.125.06 is also tested with all the experiments while not included in this paper). On the software side, Python3.9, CUDA 11.7 , cuDNN 8.8.0, PyTorch 2.0.1, and Torchvision 0.15.2 are used. These configurations constitute the hardware and software environment of our experiments, ensuring the consistency and reproducibility of our experiments.

\textbf{Models} Our tests mainly considered popular DNNs models as shown in Table~\ref{table:models}.

\subsection{Experimental scheme I, "-rdynamic" vs "base"}
\label{sec:rdynamicvsbase}

\begin{figure}[tp]
	\centering
	\includegraphics[width=\columnwidth]{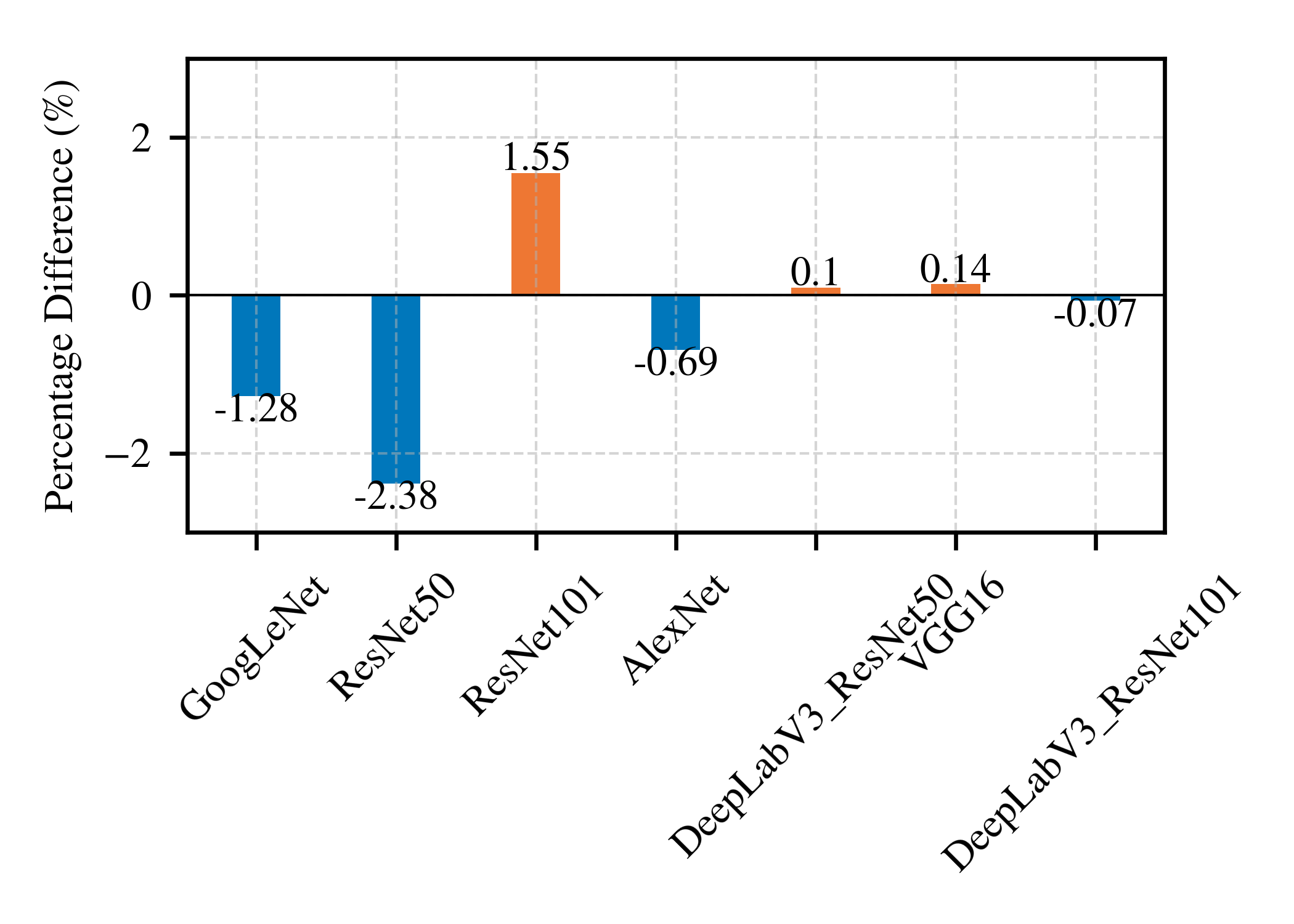}
    \caption{Percentage Difference in JCT (rdynamic vs base)}
    \label{fig:rdynamicvsbase}
\end{figure}
"-rdynamic" environment refers to the PyTorch 2.0.1 and Torchvision 0.15.2 version recompiled with the "-rdynamic" option, and the "base" environment refers to the PyTorch 2.0.1 with the default compilation option and Torchvision version 0.15.2. Running inference services based on different models in the "-rdynamic" and "base" environments respectively. Each service runs 1000 inference tasks, and then we measures the average JCT.

We tested seven groups of common models, as shown in Fig~\ref{fig:rdynamicvsbase}. The difference in JCT between the two environments is very small, within -2.38\% to 1.55\%, with GoogLeNet, ResNet50, AlexNet, and deeplabv3\_resnet101 running faster in the "-rdynamic" environment than in the "base" environment. The other models, however, are slower. Theoretically, "the -rdynamic option instructs the linker to add symbols to the symbol tables that are not normally needed at run time. Specifically, since symbol table lookups in GNU based systems are based on hash tables, having more symbols increases the chance that there would be hash collisions. Having more symbols collide in the hash means it will take more time to resolve each dynamic symbol."\cite{stackoverflow} This performance difference can be ignored if the symbol table of the c library called by pytorch/tensorflow is not very large.

Therefore, from the comprehensive analysis of test results and theory, the subtle differences in performance shown in Fig~\ref{fig:rdynamicvsbase} are more likely to be measurement errors, and we can assume that the effect of recompiling with the "-rdynamic" option on program operation is almost negligible.

\subsection{Experimental Scheme II, Single ML Inference Service FIKIT vs Base}
\label{sec:schemeFIKITVSbase}
The "single ML Inference services" environement refers to the setting that we are only hosting one inference service' model on a dedicated GPU. The model has already been profiled and entered long-term "sharing mode" that is ready to share the device with other services. The NVIDIA default mode is the same as "base" environement described in ~\ref{sec:rdynamicvsbase}. It is a common and basic scenario that can evaluate the overhead of applying FIKIT architecture compared to NVIDIA default mode but not invovling more complicated multi-task sharing scenario which will be evaluated in ~\ref{sec:schemeMultitask}.

\begin{figure}[tp]
	\centering
	\includegraphics[width=\columnwidth]{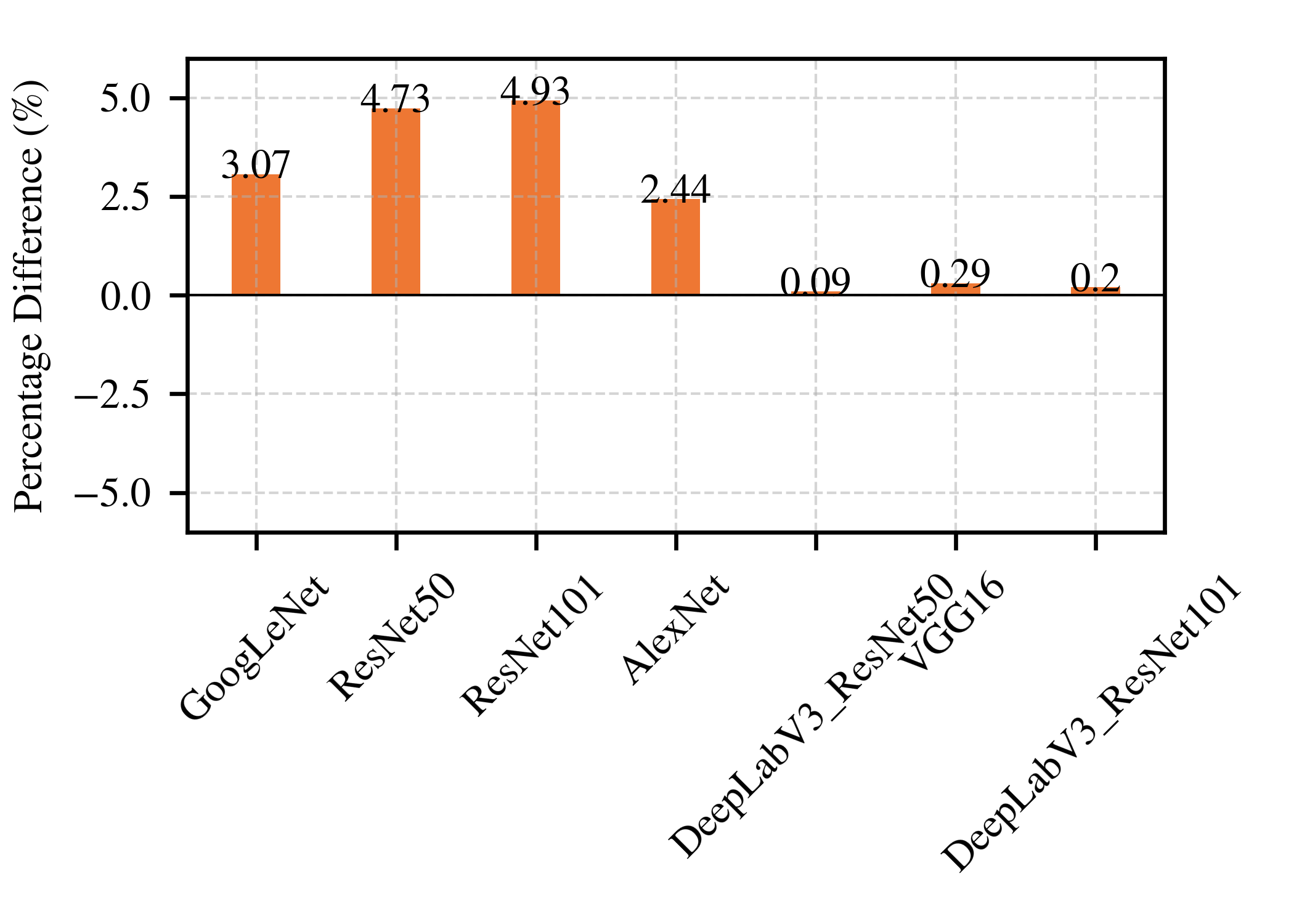}
    \caption{Single Service JCT Ovearhead FIKIT Sharing Stage vs Nvidia Default Mode}
    \label{fig:singleFIKITVSbase}
\end{figure}

We run inference services based on different models in the FIKIT and base environments respectively. Each service runs 1000 inference tasks, and then we measures the average JCT. The result measured models' JCT percentage increase is shown in Fig~\ref{fig:singleFIKITVSbase}. 

For the seven groups of common models, the percentage increase of JCT in FIKIT environement compared to the base environement is small, within 0.09\% to 4.93\%. Our current implementation of the FIKIT architecture managed to limit the long-term runing service overhead under 5\% compared to the default NVIDIA environement, making it practical to be adopted in a cloud environement.

\subsection{Experimental Scheme III, Single ML Inference Service FIKIT Sharing Stage vs Measuring Stage}
We have evaluated a profiled ML inference model's overhead compared to the base environement. This experiment takes a look at the overhead of hosting a new ML inference service, referred as the measuring stage. It is a temporary stage for an unseen service using FIKIT architecture. 

The "single ML Inference service" setting is the same as in Scheme II, section ~\ref{sec:schemeFIKITVSbase}. The "FIKIT measuring stage" refers to the stage where a new service's task is executed with kernel-level measurement and described in section 3.2. The "FIKIT sharing stage" is the same FIKIT mode used in Scheme II, section ~\ref{sec:schemeFIKITVSbase} and all the other experiment's FIKIT mode.

\begin{figure}[tp]
	\centering
	\includegraphics[width=\columnwidth]{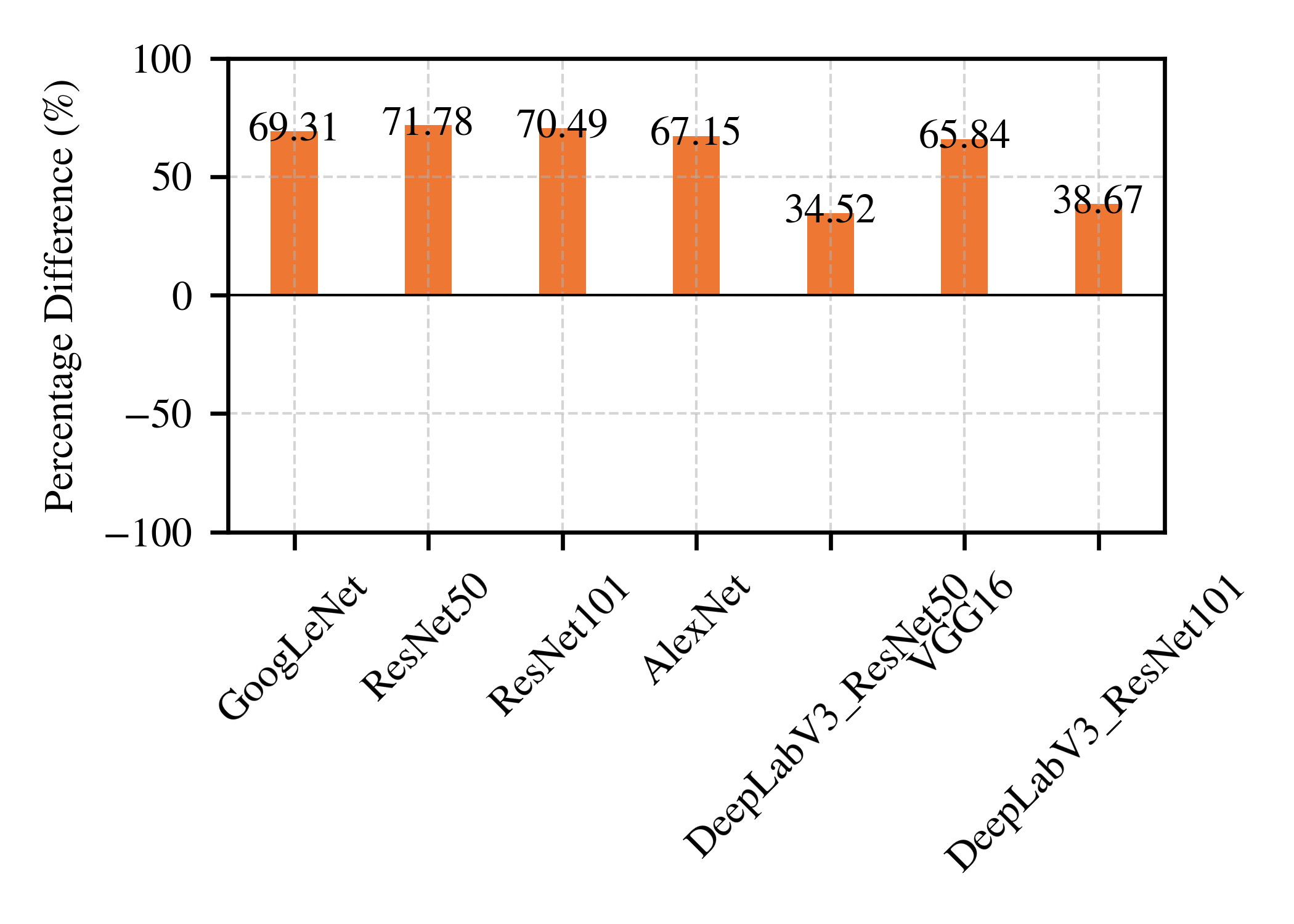}
    \caption{Single Service JCT Ovearhead FIKIT Measuring Stage vs Nvidia Default Mode}
    \label{fig:singleMeasureVSbase}
\end{figure}

We run inference services based on different models in the FIKIT measuring and Nvidia default environement respectively. Each service runs 1000 inference tasks, and then we measures the average JCT. The result models' JCT percentage increase is shown in Fig~\ref{fig:singleMeasureVSbase}. 

For the seven groups of common models, the percentage increase of JCT in FIKIT measuring stage compared to the base environement is large: a 34.52\% up to 71.78\% additional time to complete a task. This overhead would offset the benefits brought by kernel-level scheduling if an online measuring stage is applied in the real-time sharing scenario.

Compared to the 0.09\% to 4.93\% overhead of FIKIT sharing mode in Fig~\ref{fig:singleFIKITVSbase}, this experiment's result demonstrate the necessity and effectiveness of dividing the FIKIT scheudling system into measuring and sharing stages. We isolates the cost of kernel-level profilation from the kernel-level real-time scheduling in the long-run with the kernel identical technique. This design results the overall FIKIT overhead under 5\% for repeating cloud services. 

\subsection{Experimental Scheme IV, Multiple ML Inference Services Sharing}
\label{sec:schemeMultitask}
In a multitasking and containerized cloud computing environment, it is a common scenario for multiple inference services' tasks to run concurrently in the same GPU instance. We compared the performance of FIKIT sharing mode with default GPU sharing and exclusive mode respectively.

\subsubsection{JCT Comparison Test (FIKIT VS Share)}
We combined services running different models and measure their tasks' JCT values in FIKIT mode and default sharing mode respectively, and test each combination 1000 times. The task completion time of different models is different. For straight-forward comparison, we take the tasks execution records of the two services overlapping fully. For example, taking two services concurrently running as an example, service A issues keypointrcnn\_resnet50\_fpn model inference tasks, and service B issues fcn\_resnet50 inference tasks. Both services perform 1000 inferences concurrently in FIKIT mode and Share mode respectively.  The total time of each service is shown in the table~\ref{table:JCTComparisonTest}. From table~\ref{table:JCTComparisonTest}, it can be concluded that from the beginning of the two services to the 16th second, there are always two concurrent tasks from different services running on the GPU for both experiment settings.

\begin{table}[h!]
    \centering
    \begin{tabular}{|c|c|c|}
    \hline
    \multicolumn{3}{|c|}{1000 total execution time (seconds)} \\
    \hline
    & Service A & Service B \\
    \hline
    Default GPU Sharing Mode & 38.15803 & 16.02363\\
    \hline
    FIKIT mode & 33.12627 & 39.09784\\
    \hline
    \end{tabular}
    \caption{Total Execution time for A's Tasks and B's Tasks under Two Modes}
    \label{table:JCTComparisonTest}
\end{table}

\begin{figure}[tp]
	\centering
	\includegraphics[width=\columnwidth]{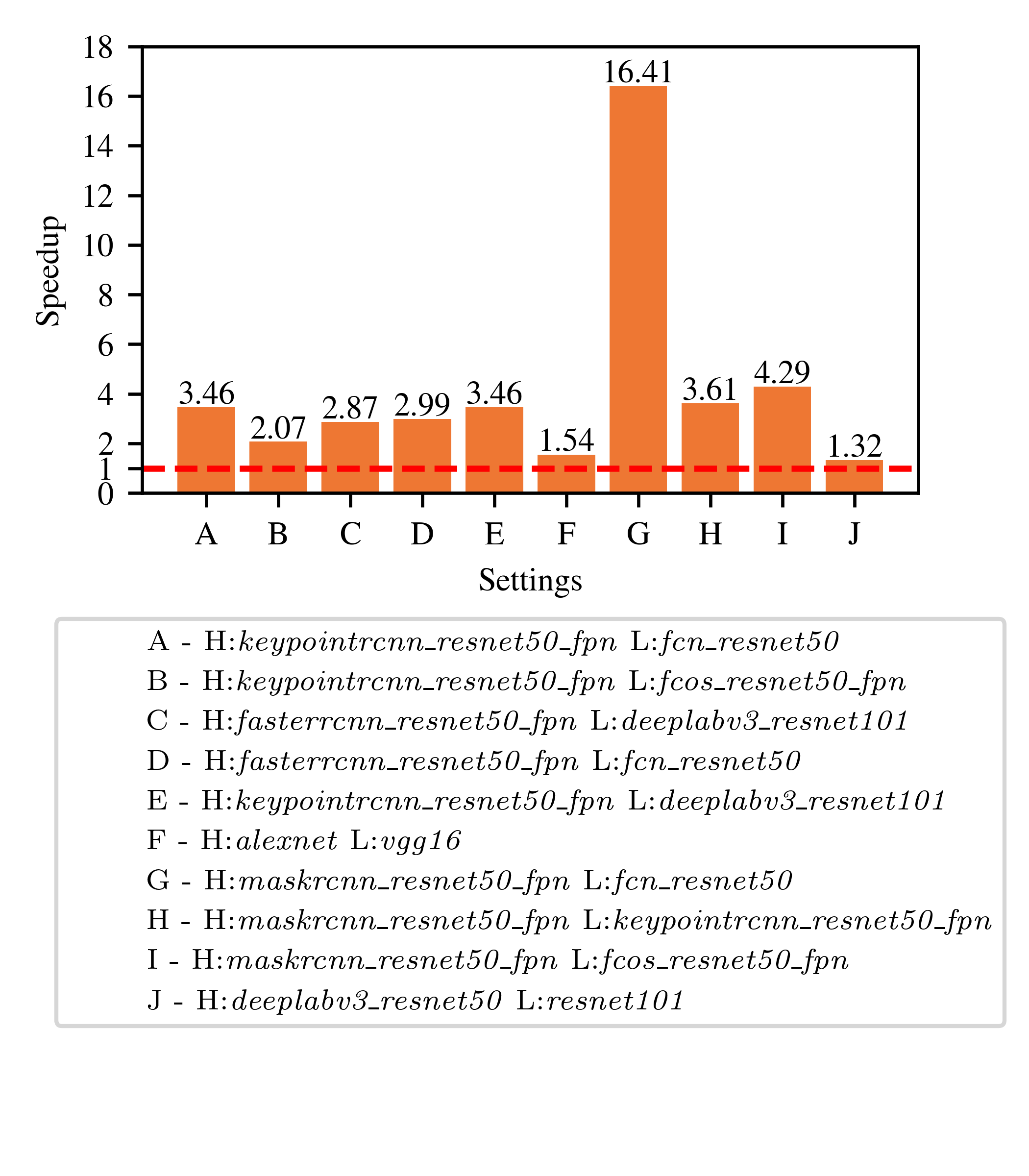}
    \caption{High Priority JCT Speedup of FIKIT over Default GPU Sharing mode}
    \label{fig:highPrioJCTSpeedupFIKITShare}
\end{figure}

Therefore, for this services combination, only the first 16 seconds of JCT data were collected for evaluation. Using the same test method above, 10 combinations of services were tested, and the test results are shown in Fig~\ref{fig:highPrioJCTSpeedupFIKITShare} where "H" indicates that the model in this mode is a high-priority task, and "L" indicates that the model in this mode is a low-priority task. In the test cases of 10 services combinations, FIKIT's JCT is accelerated by 1.32 to 16.41 times compared to the JCT in GPU sharing mode, and more than half of the cases are accelerated by more than 3.4 times, which is a significant performance improvement. If we need both concurrent services serving and guaranteed priority response for some tasks, FIKIT is a good choice.

\begin{figure}[tp]
	\centering
	\includegraphics[width=\columnwidth]{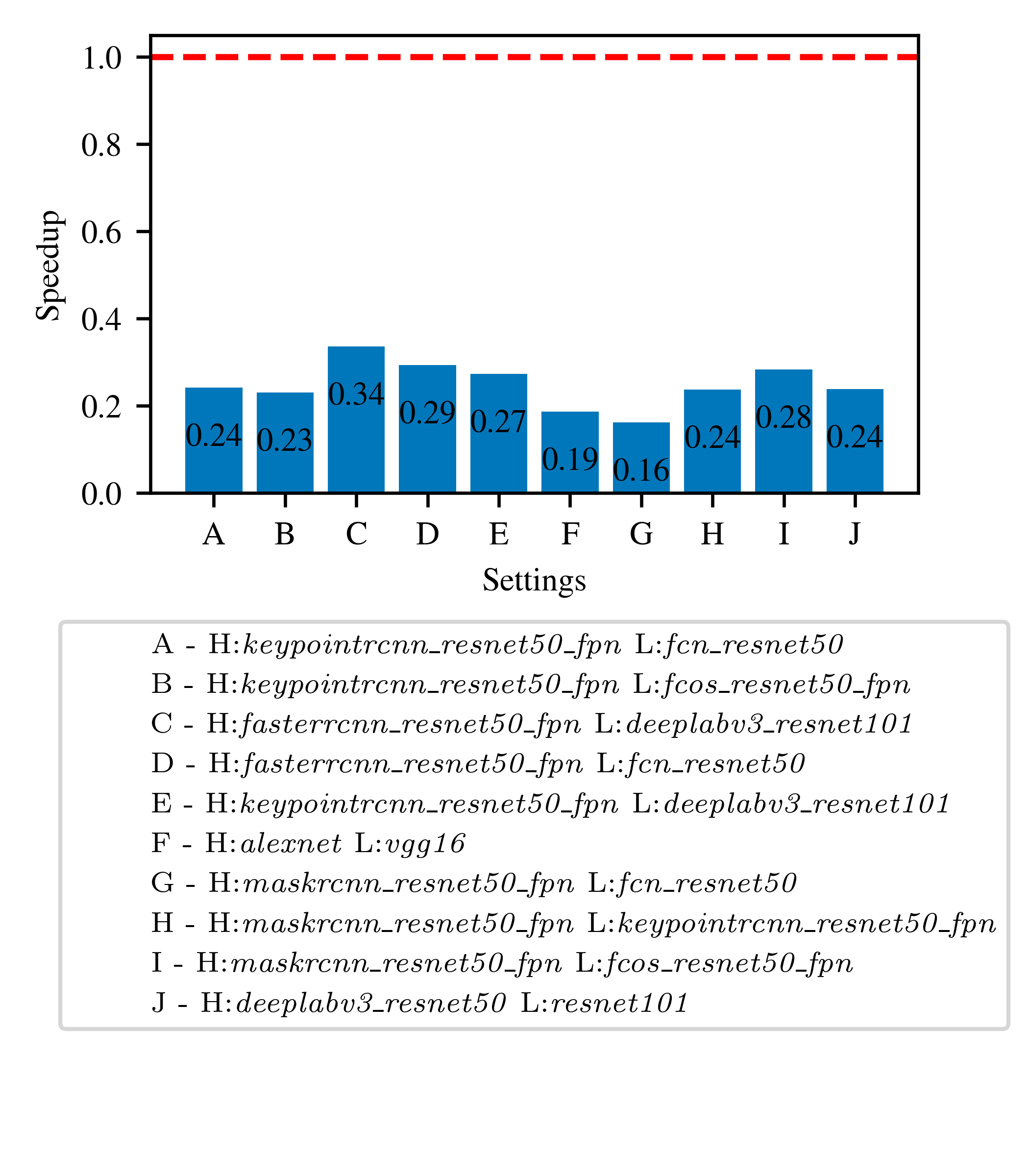}
    \caption{Low Priority JCT Speedup of FIKIT over Default GPU Sharing mode}
    \label{fig:lowPrioJCTSpeedupFIKITShare}
\end{figure}

By design, FIKIT scheduling algorithm will give priority to service A's high-priority tasks. The default GPU sharing mode, however, does not distinguish tasks priorities, and service A's tasks are mostly delayed. Of course, FIKIT's schedule also needs to pay a price, when tasks from A is executed first, tasks issued by B will inevitably be delayed for a longer period. As shown in Fig~\ref{fig:lowPrioJCTSpeedupFIKITShare}, the operation efficiency of B's tasks in most combinations is less than 30\% of that in share mode in FIKIT mode. This is because the FIKIT model focuses on prioritizing high-priority tasks.

\subsubsection{Low Priority JCT Comparison Test (FIKIT VS Exclusive)}
\begin{figure}[tp]
    \centering
	\includegraphics[width=\columnwidth]{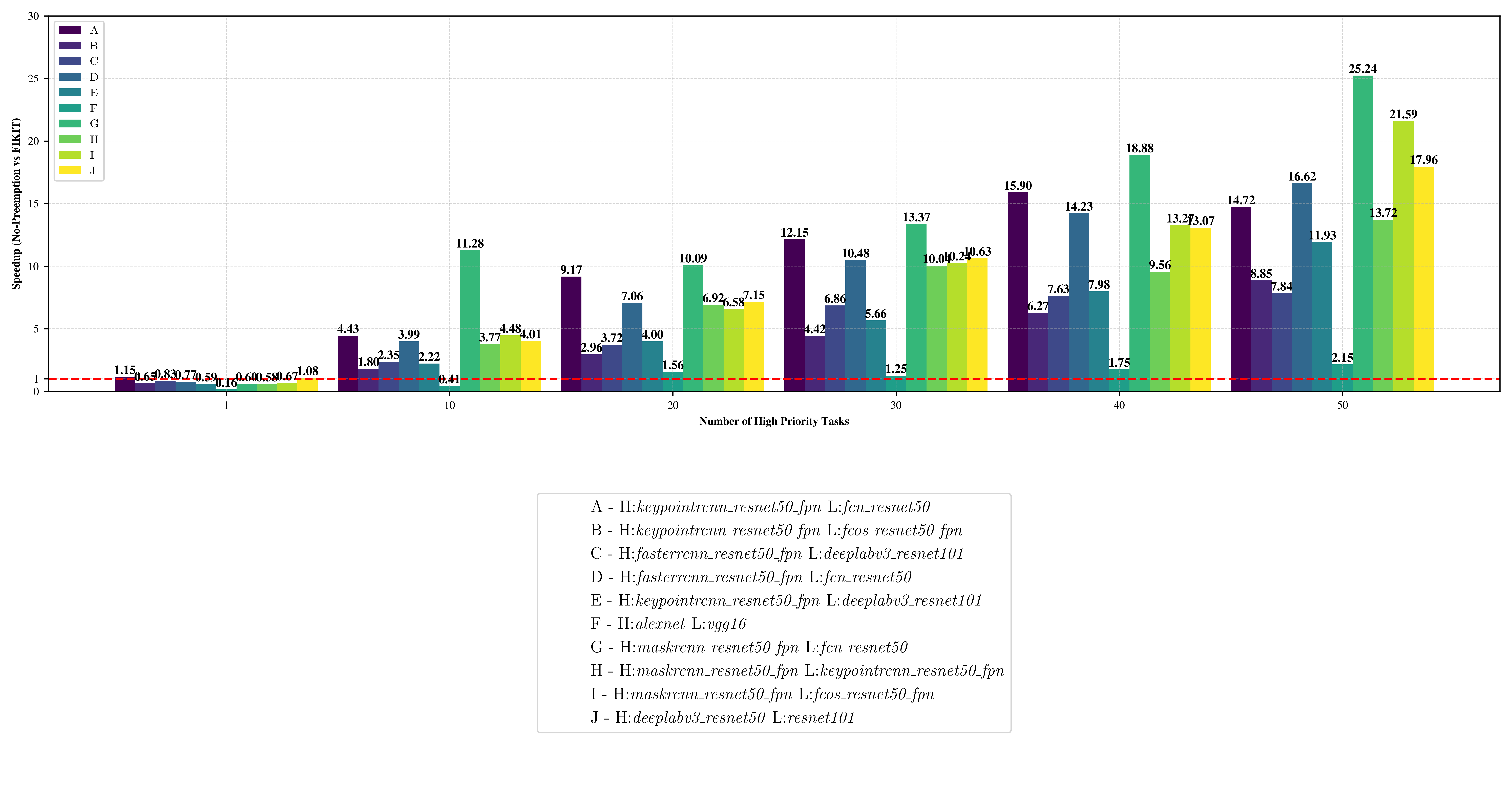}
    \caption{Low Priority JCT Speedup of FIKIT over Exclusive mode}
    \label{fig:lowPrioJCTSpeedupFIKITExclusive}
\end{figure}

When service A issued more or much more tasks than the low-priority service B, we concurrently run services of different models and compare the low priority tasks' JCT values in FIKIT mode and Exclusive mode. Since the GPU's Exclusive mode cannot run two tasks at the same time, we execute the two services sequentially, measure their execution times separately, and then calculate their JCT values if they are requesting a GPU at the same time. To demonstrate how low priority tasks are delayed, the ratio of A's tasks to B's tasks in the experiment are 1:1, 10:1, 20:1, 30:1, 40:1, and 50:1. As shown in Fig~\ref{fig:lowPrioJCTSpeedupFIKITExclusive}, when 1:1 tasks from both services are executed, low priority tasks' JCT under exclusive mode is close to that of the FIKIT mode. From 10:1 to 50:1, the ratio of JCT in exclusive mode to JCT in FIKIT mode, however, shows a linear upward trend, showing low priority task's JCT getting more and more delays under exclusive mode while FIKIT mode's keep constant. Since the two services' tasks need to run in order of priorities, the JCT of B's tasks in exclusive mode is the sum of the execution time of itself and the time waiting for the completion of A's tasks. If there are more tasks with high priority, the JCT of the low-priority task will be longer under exclusive mode. This is the main disadvantage of using the exclusive mode. In our experiment, we also found that when we set the GPU to exclusive mode, we can't run two "cloud services" at the same time, and the later service will fail because it can't occupy the GPU. Therefore, GPU exclusive mode is not suitable for multi-tasking and containerized cloud computing environments, both in terms of operation and performance.


\subsubsection{High Priority JCT Comparison Test in Preemption Scenario (FIKIT VS Share)}
\begin{figure}[tp]
	\centering
	\includegraphics[width=\columnwidth]{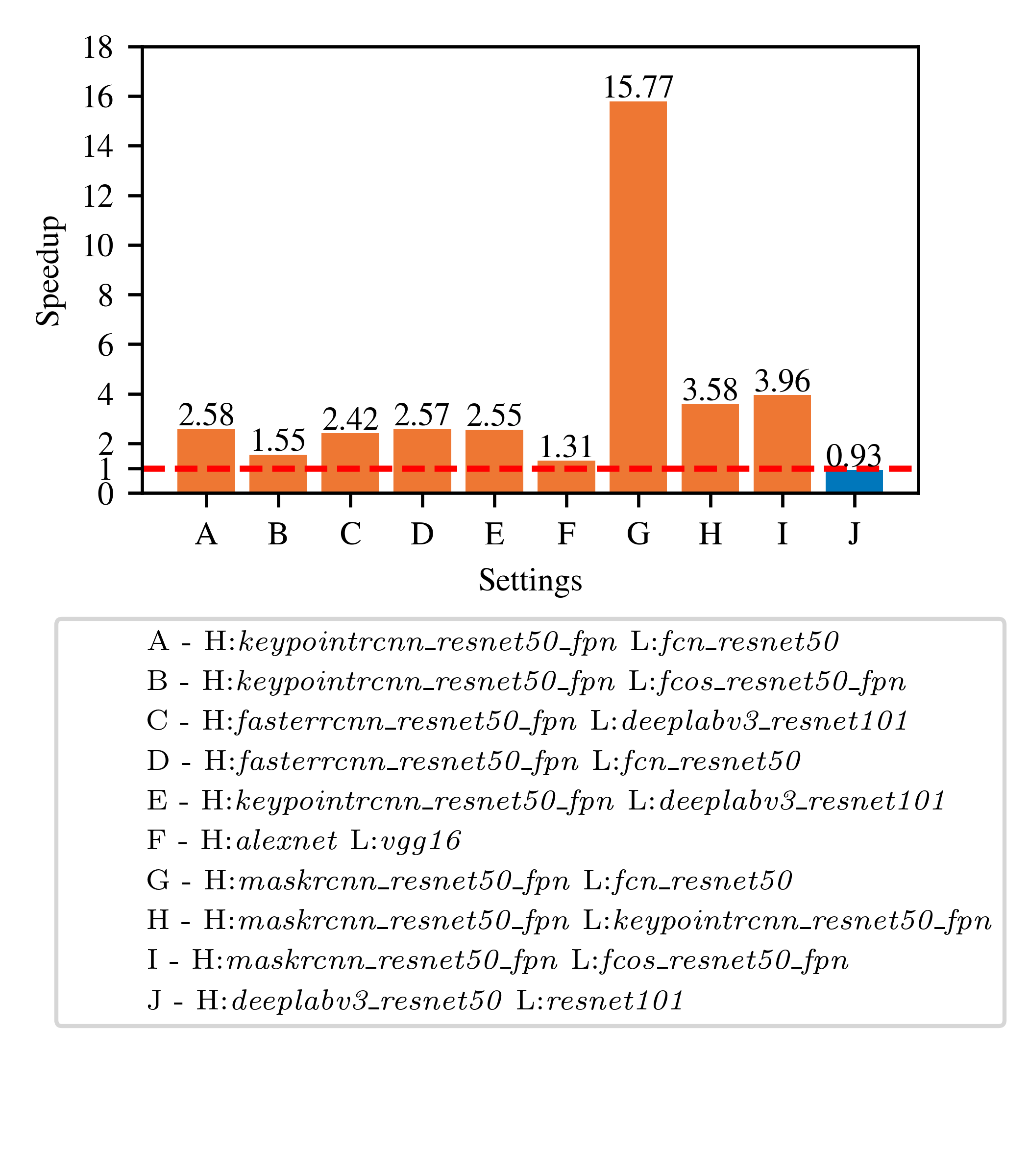}
    \caption{High Priority JCT Speedup of FIKIT VS Default GPU Sharing mode w Preemption}
    \label{fig:highPrioJCTSpeedupFIKITSharePreemptive}
\end{figure}

The setting for this test is that service B's low-priority tasks run continuously, and the service A's high priority tasks executes intermittently. When B's low-priority task runs continuously, service A issues a high-priority task every 1 second, and there are in total 100 such tasks. We measure the JCT value of A's tasks in FIKIT mode and default GPU sharing mode and calculate the average JCT value of 100 tasks.  As shown in Fig~\ref{fig:highPrioJCTSpeedupFIKITSharePreemptive}, under most models' combinations, the JCT of A's tasks preempting the running GPU tasks under the FIKIT mode is significantly faster than the JCT under the default GPU sharing mode, up to 15.77 times speedup. However, for deeplabv3\_resnet50 and resnet101 combination, the FIKIT mode high-priority tasks' JCT increased compared to the default sharing mode.

\begin{figure}[tp]
	\centering
	\includegraphics[width=\columnwidth]{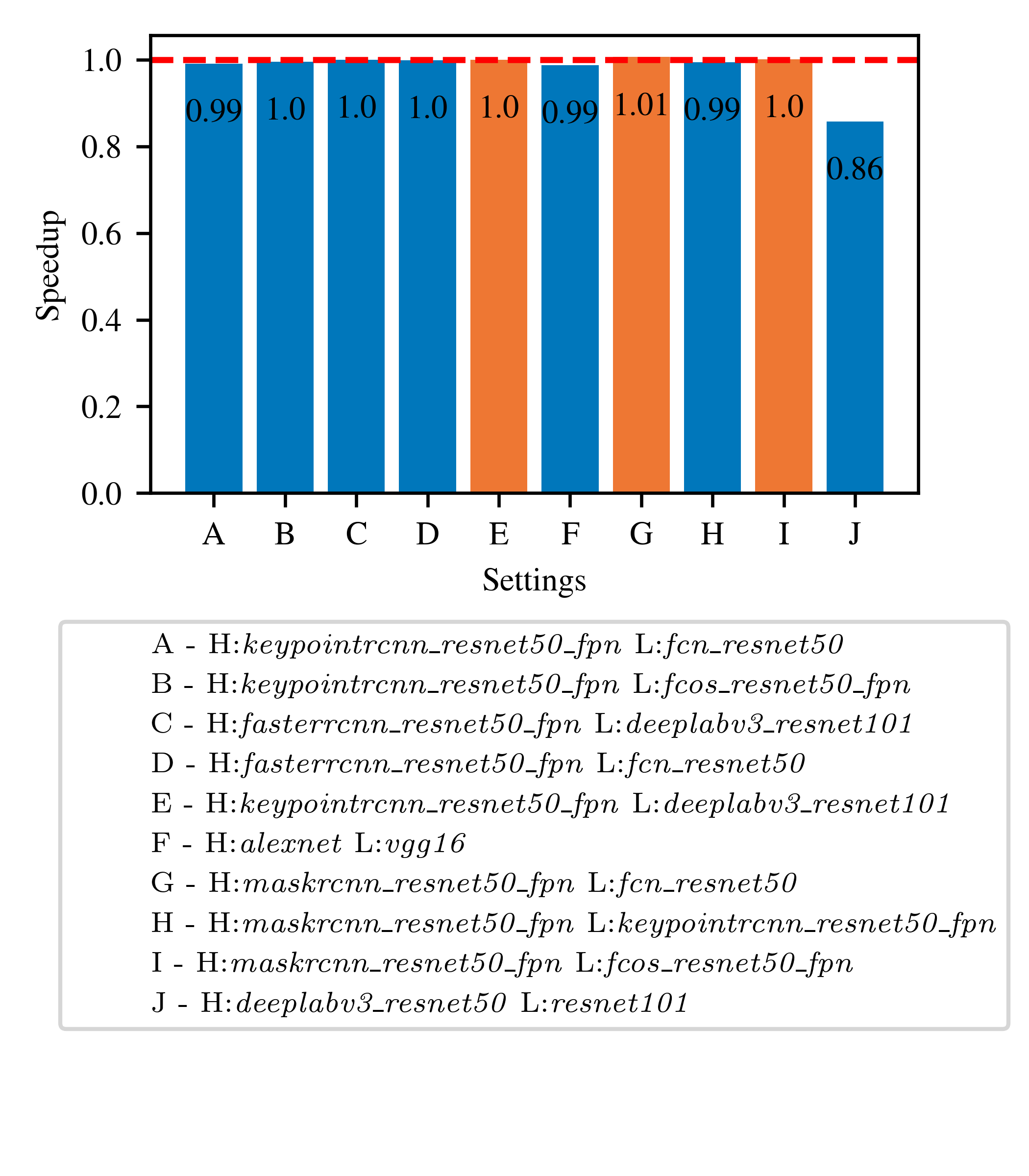}
    \caption{Low Priority JCT Speedup of FIKIT VS Exclusive mode w Preemption}
    \label{fig:lowPrioJCTSpeedupFIKITSharePreemptive}
\end{figure}

As shown in Fig~\ref{fig:lowPrioJCTSpeedupFIKITSharePreemptive}, the difference between the JCT values of B's low-priority tasks in FIKIT mode and default GPU sharing mode is very small, except for the ratio of deeplabv3\_resnet50 to resnet101 concurrent tasks, which is only 0.86. The other model combinations' ratios are very close to each other. Thus, the effect of the FIKIT mode on the operation of service B is almost negligible in this scenario.

From this test, we can see that not all combinations of ML model services using FIKIT scheduling mode will be better than the default GPU sharing mode, for example: deeplabv3\_resnet50 and resnet101 run concurrently on the same GPU instance. Therefore, when using FIKIT scheduling, we need to choose a good combination of GPU sharing services to achieve the best concurrency efficiency. FIKIT scheduling has obvious advantages in the multi-task and containerized cloud computing environment because there are non-stopped computation requests come with different priorities, having non-symmetric impact on QoS for sharing a GPU device. We schedule and select a good combination of tasks to run concurrently on the same GPU, to achieve the best operation efficiency. Of course, what kind of combination of tasks will achieve the best results also needs further measurement and research in the future.

\subsubsection{Low Priority JCT Stability Test in FIKIT Sharing Scenario}
\begin{figure*}[tp]
    \centering
    \includegraphics[width=\textwidth]{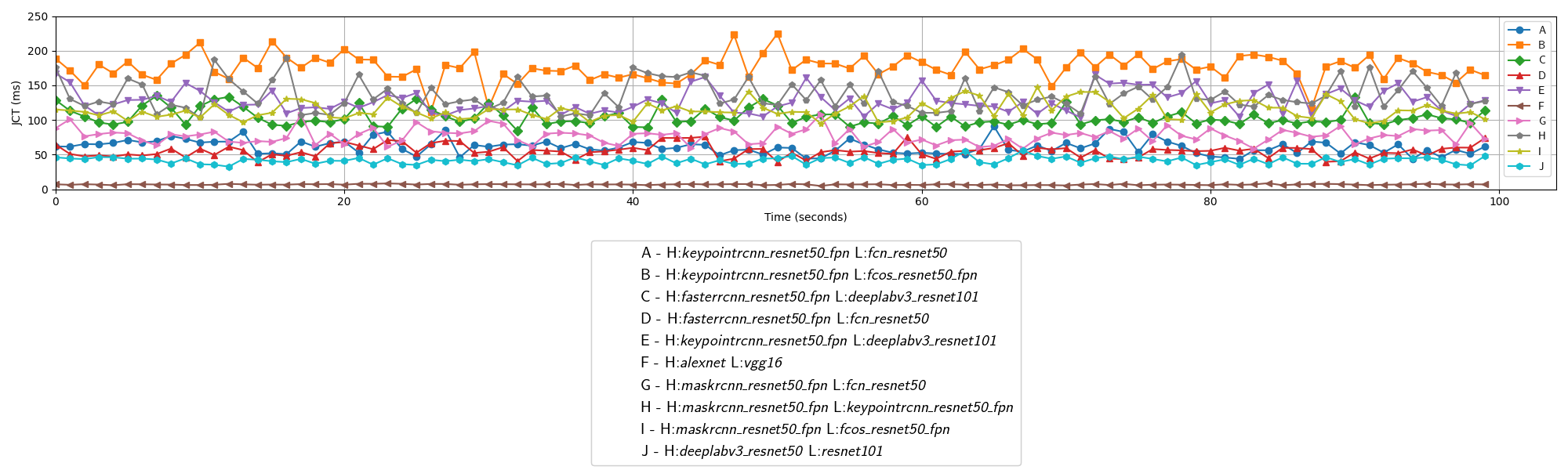}
    \caption{Low Priority JCT time lines under FIKIT sharing}
    \label{fig:lowPrioJCTTimelines}
\end{figure*}

The setting for this test is that service A runs high-priority tasks continuously, and service B executes low-priority tasks intermittently. In detail, service B issues a low-priority task every 1 second, and there are in total 100 such tasks. We measure the JCT value of B's tasks in FIKIT mode and calculate the coefficient of variation to evaluate low-priority tasks JCT stability spanning 100 seconds.  As shown in Fig~\ref{fig:lowPrioJCTTimelines}, under every models' combinations, the JCT of low-priority tasks scavenging the inter-kenel idling time of high-priority GPU tasks under the FIKIT sharing mode is stable and predictable. We also show that coefficient of variations are within 0.095 to 0.164 for 10 combinations of models spanning 100 seconds in Table~\ref{table:stabilityTimelines}. These timelines' CV being way smaller than 1 indicates a low variability and high predictability of low-priority tasks' job completion time when there are continuously runing high-priority tasks using FIKIT sharing mode.

The stability and predictable JCT for low-priority tasks is a desired guarantee in cloud computing environement and we show that FIKIT have achieved this guarantee under multi-task GPU sharing scenario.

\begin{table}[h!]
    \centering
    \begin{tabular}{|c|c|c|c|}
    \hline
    \textbf{Timeline} & \textbf{$\sigma$ (ms)} & \textbf{$\mu$ (ms)} & \textbf{CV = $\frac{\sigma}{\mu}$} \\
    \hline
    A & 10.047 & 61.391 & 0.163657 \\
    B & 18.428 & 176.578 & 0.104361 \\
    C & 11.948 & 104.195 & 0.114673 \\
    D & 8.692 & 55.408 & 0.156871 \\
    E & 15.962 & 127.269 & 0.125423 \\
    F & 0.670 & 7.017 & 0.095457 \\
    G & 10.524 & 77.409 & 0.135951 \\
    H & 21.547 & 136.410 & 0.157956 \\
    I & 12.393 & 114.532 & 0.108204 \\
    J & 4.616 & 41.719 & 0.110651 \\
    \hline
    \end{tabular}
    \caption{Standard Deviation, Mean, and Coefficient of Variance for each Timeline}
    \label{table:stabilityTimelines}
\end{table}

\section{Discussion and Future Work}
This section will cover how to make the FIKIT model work best at runtime, and where future research based on FIKIT can work on. 

\text{Cluster-level GPU Tasks Scheduling:} the FIKIT algorithm is an algorithm to orchestrate concurrent tasks loaded on a single GPU instance. In the cloud computing environment, we also need to implement a cluster-level scheduling policy to decides which concurrent tasks should be allocated to share the same GPU device, meeting a GPU instance's resources cap, and then at the device-level scheduling these tasks' kernels execution through the FIKIT algorithm. The design and implementation of this cluster-level scheduling policy is one of the goals that need to be studied in the future.

\textbf{What Tasks are Suitable for Sharing a GPU:} The FIKIT algorithm utilizes inter-kernel GPU idle time within high-priority tasks to execute low-priority tasks' GPU kernels. Given a fix amount of GPU uptime, more kernel from different priority level can be completed while minimizing their impact on executing the high-priority tasks. It has been shown from the evaluations that different tasks running various models exhibits largely different kernel execution time and inter-kernel idle time. Thus, different combinations of concurrent tasks in FIKIT mode may lead to a large difference in execution efficiency of the low priority tasks scavenging inter-kernel idle times. For example: The combination of maskrcnn\_reset50\_fpn and fcn\_reset50 works well in the 4.3.3 scenario, but deeplabv3\_resnet50 is largely impact when concurrent running with resnet101. In our tests, we found that this problem also exists in Nvidia's default sharing mode. Therefore, to have the best GPU concurrency with FIKIT model, further research is needed on what kinds of tasks are optimal for sharing a GPU together.

We can prepare combinations of potential models and measure their enhancement and impact in their JCT when sharing on the same device. These measurements will be preloaded for prediction in an cluster-level scheduling policy. When a task request arrives, the policy finds the GPU on which its optimal matching task resides using the preloaded measurement data, and then schedules it to concurrently run on that GPU.

We can also try to find the matching pattern of tasks combinations by analyzing "good" combinations or "poor" combinations, convert observations into an algorithm and implement it in the cluster-level scheduling policy. Of course, this is a non-trivial work.

\textbf{Exclusive Mode GPU Sharing:} exclusive executing tasks is not friendly in the cloud computing environment at this point because default exclusive mode does not allow multiple tasks to occupy the same GPU, resulting in the inability to allocate multiple cloud services using a GPU together, even if they are not invoked to run. We can implement a new scheduling mode based on FIKIT, which would support allocating multiple GPU services on one GPU but exclusively run one task at a time, achieving a software defined GPU exclusive mode.

\section{Related Work}
\textbf{GPU sharing and kernel preemption.} Nvidia vGPU\cite{5} virtualizes the GPU into multiple logical units for time-divisional multiplexing among virtual machines. They address the challenges of GPU sharing and scheduling among virtual machines, but the lack of preemptive mechanisms limits efficiency and flexibility. The NVIDIA device plugin for Kubernetes\cite{8} presents a GPU isolation solution in containerized environments, allowing elastic access to GPUs at a fine-grained level. However, it is unable to preempt low-priority tasks. Nvidia MPS\cite{4} enables GPU scheduling and sharing at the process level. However, MPS does not support priority preemption. TimeGraph\cite{10} achieves fine-grained scheduling through GPU hardware preemption support. It is specifically designed and implemented for Nouveau\cite{9}, relying on open-source GPU driver. Gemini\cite{13} and KubeShare\cite{14} are two user-space runtime scheduling frameworks that both achieved fine-grained GPU allocation control through support for multi-tenancy and elastic provisioning. They are again a non-preemptive, GPU fair sharing approach. REEF\cite{30} is the GPU-accelerated DNN inference serving system that enables microsecond-scale kernel preemption and controlled concurrent execution for GPU scheduling. It allows launching real-time kernels by proactively killing and restoring "best-effort kernels", and utilizes GPUs fully by dynamically padding real-time kernels with appropriate best-effort ones, incurring negligible overhead. Its implementation relies on AMD ROCm\cite{42}(an open-source GPU computing platform) to enable kernel scheduling; The scheduling mechanism of REEF for kernels differs significantly from CUDA and cannot be applied in closed-source CUDA environment, thus it only supports one AMD Radeon Instinct MI50 GPU model. Share-a-GPU\cite{35} provides an abstraction where, in a round-robin manner, every workload can use a GPU(s) over a time quantum exclusively. It focuses on resource allocation and does not study kernel priority and preemption. Preemption of a CUDA Kernel Function\cite{38} implements application-level checkpointing for CUDA kernels to enable preemption of kernel execution, but this checkpointing method has a relatively high overhead. Kernelet\cite{39} proposes to improve GPU kernel throughput. It identifies slice points automatically and generates slicing schemes at runtime considering data locality. A two-level scheduler coordinates concurrent execution of fine-grained kernel slices. However, the approach relies on compile-time analysis and runtime profiling which may limit adaptability. Dynamic slicing scheduling incurs runtime overheads that can become bottlenecks for complex kernels. FLEP\cite{17}'s preemption capability can help improve the system's average responsiveness by preempting long-running kernels to reduce the waiting time for short-running kernels.

\textbf{Measurement of kernel execution time.} The CUDA Flux profiler\cite{40} uses code instrumentation to keep track of how often threads execute a specific basic block, recording the basic block execution frequencies and PTX(Parallel Thread execution) instruction counts. Braun et al.\cite{22} obtained relevant Kernel Metrics using the CUDA Flux profiler to estimate the kernel execution time. The code instrumentation method requires recompiling the client source code and may significantly impact efficiency. FLEP \cite{17} follow the performance modeling methodology used by Chen et al. \cite{37} and build lightweight kernel-specific models through linear regression. Based on this model, FLEP calculates the kernel execution time online. The performance modeling methodology is a method that has a relatively large measurement error, which can be further amplified when the program is run on different GPU hardware. Likewise, Gemini\cite{13} and KubeShare\cite{14} also employs cudaEventSynchronize to log the kernel execution time, but it utilizes an online approach for measurement, which tends to substantially impact the performance of the program. GPUShare\cite{15} profiles running times of kernels on the GPU in order to predict and track running times of the same kernels when they are issued again and to distinguish short- and long- running kernels. However, it does not provide specific details about the profiling methods, nor does it explain how to determine if kernels are the same as previously executed kernels. REEF\cite{30} developed a kernel profiler within the open-source AMD GPUOpen code to measure the execution time of each kernel, which lacks generality and cannot be applied to CUDA programs.

\section{Conclusion}
Aiming at cloud computing environment, this paper proposed a novel GPU kernel-level preemptive multi-task scheduling architecture, the FIKIT: Filling Inter-kernel Idle Time. Instead of eliminating inter-kernel GPU idle time, we fully utilize these gaps of a time-sensitive GPU task to concurrently execute other sharing GPU tasks, serving multiple tasks having non-symmetric impact on QoS when sharing a GPU device. By a novel mechanism to identify GPU kernels at runtime, the FIKIT find an efficient way to apply offline fine-grained kernel profiling during runtime scheduling with feedback mechanism, separating the profiling impact from GPU multi-task sharing. Across a set of ML models, the FIKIT based inference system accelerated high priority tasks by 1.32 to 16.41 times compared to the job completion time(JCT) in default GPU sharing mode, and more than half of the cases are accelerated by more than 3.4 times. Alternatively, under preemptive sharing, the low-priority tasks have a comparable to default GPU sharing mode JCT, with a 0.86 to 1 times ratio.

In additon, this work discovered an efficient GPU kernel identification mechanism, which retrieves a  kernel's function name with no additional cost at runtime and is compatible with closed-source GPU drivers. This new mechanism enables offline kernel measuring statistics to be connected with kernels identified at real-time, solving the real-time kernel-level scheudling overhead problem of this work: an overhead under 5\%. and we believe it can be applied to solve broader GPU related problems.

\bibliographystyle{ACM-Reference-Format}
\bibliography{reference}

\end{sloppypar}
\end{document}